\documentclass[final,5p,times,twocolumn,authoryear]{elsarticle}

\graphicspath{{figs1/}}
\usepackage{amssymb}
\usepackage{amsmath}
\usepackage{graphicx}
\usepackage{subcaption}
\usepackage[labelfont=bf]{caption}
\usepackage{float}
\usepackage{hyperref}
\hypersetup{
    colorlinks=true,
    linkcolor=blue,
    filecolor=magenta,      
    urlcolor=blue,
}
 \usepackage{cleveref}
\usepackage{placeins}
\usepackage{bm,csquotes}
\usepackage{enumerate}
\usepackage{booktabs}
\usepackage[ruled,vlined]{algorithm2e}
\SetKwComment{Comment}{/* }{ */}
\RestyleAlgo{ruled}

\usepackage{siunitx}
\usepackage{multirow}
\usepackage{adjustbox}
\usepackage{tikz}
\usetikzlibrary{arrows.meta, positioning, shapes.geometric}

\makeatletter
\renewcommand\paragraph{\@startsection{paragraph}{4}{\z@}%
  {1ex \@plus 1ex \@minus .2ex}%
  {-1em}%
  {\normalfont\normalsize\bfseries}}
\makeatother

\journal{Artificial Intelligence In Medicine}
\begin{document}

\begin{frontmatter}



\title{Bayesian Inference of Nonlinear Malaria Dynamics in Ghana via an Ensemble Markov Chain Monte Carlo Sampler}

%%%  HERE
\author[label2]{T. Ansah-Narh\corref{cor1}}
\cortext[cor1]{Corresponding author.}
\ead{theophilus.ansah-narh@gaec.gov.gh}

\author[label1]{Y. Asare Afrane} %% Author name
\ead{yafrane@ug.edu.gh}

\author[label2]{J. Bremang Tandoh} %% Author name
\ead{joseph.tandoh@gaec.gov.gh}
%%%  HERE
% \author[label5]{R. Adjetey Laryea}
% \ead{ravenhill.laryea@upsamail.edu.gh}
%%%%%%%%%%%%%%%%%%%%%%%%%%%%%H1%%%%%%%%%%%%%%%%%%%


% % %% Author affiliation
\affiliation[label2]{organization={Ghana Space Science and Technology Institute, Ghana Atomic Energy Commission},%Department and Organization
            addressline={P. O. Box LG 80}, 
            city={Legon},
            % postcode={00233}, 
            state={Accra},
            country={Ghana}}
            
\affiliation[label1]{organization={Department of Medical Microbiology, University of Ghana Medical School}, 
            % addressline={P. O. Box 4236}, 
            city={University of Ghana},
            % postcode={00233}, 
            state={Accra},
            country={Ghana}}

% \affiliation[label2]{organization={Department of Statistics and Actuarial Science, University of Ghana},%Department and Organization
%             addressline={P. O. Box LG 115}, 
%             city={Legon},
%             % postcode={00233}, 
%             state={Accra},
%             country={Ghana}}

% \affiliation[label1]{organization={Department of Information Technology and Mathematical Sciences, Methodist University Ghana},%Department and Organization
%             addressline={P. O. Box DC 940}, 
%             city={Dansoman},
%             % postcode={00233}, 
%             state={Accra},
%             country={Ghana}}

% \affiliation[label5]{organization={ Department of Economics and Actuarial Science, University of Professional Studies, Accra},%Department and Organization
%             addressline={P. O. Box LG 149}, 
%             city={Legon},
%             % postcode={00233}, 
%             state={Accra},
%             country={Ghana}}

%% Abstract
\begin{abstract}
Reliable quantification of malaria dynamics in sub-Saharan Africa remains hindered by short, noisy, and spatially heterogeneous surveillance records that challenge the assumptions of conventional deterministic models. In Ghana, health-facility data between 2014 and 2023 reveal highly non-linear and age-specific fluctuations in hospital admissions, yet existing approaches struggle to capture such stochastic variability or provide credible uncertainty bounds. This study develops a Bayesian nonlinear inference framework that integrates a cubic deterministic baseline with a damped oscillatory kernel, estimated via an affine-invariant ensemble Markov Chain Monte Carlo sampler. The framework accommodates limited data, explicitly models parameter uncertainty, and generates probabilistic forecasts of malaria admissions for children under five years and individuals aged five years or more.
Results demonstrate that the proposed hybrid cubic–damped oscillatory kernel model achieves strong empirical adequacy ($R^{2}=0.9958$ for $<5$~years; $R^{2}=0.9956$ for $\geq5$~years) with residual errors below $2\%$ and unimodal, well-mixed posterior distributions confirming robust convergence. District-level analysis reveals pronounced spatial heterogeneity, with coefficients of variation ranging from $<0.07$ in stable urban centres such as Kumasi to $>3.3$ in peripheral districts including Mpohor and Bia~East. Forecasts for 2024--2026 indicate a gradual resurgence in admissions, increasing from approximately 137,000 to 149,000 cases among children under five and from 348,000 to 375,000 among older individuals, with uncertainty widening modestly over time.
By producing interpretable probabilistic forecasts, this Bayesian framework provides a principled tool for anticipating short-term malaria fluctuations, guiding resource allocation, and strengthening data-driven decision-making within Ghana’s national malaria control strategy.
\end{abstract}


%Research highlights
\begin{highlights}
\item Bayesian model for Ghana's malaria admissions from limited data.

\item Hybrid kernel captures smooth trends and transient fluctuations.

% \item Hybrid kernel captures smooth trends and transient fluctuations.

\item Used an affine-invariant MCMC sampler for efficient analysis.

\item Malaria admissions fell $35\%$ in under-5s from 2014--2023.

\item  Spatial variation was extreme: under $0.07$ in cities to over $3.3$ in rural areas.

\end{highlights}

%% Keywords

\end{frontmatter}



\section{Introduction}
\label{sec1}
%% Labels are used to cross-reference an item using \ref command.

Malaria remains a dominant cause of morbidity across sub-Saharan Africa and continues to impose a significant public-health and economic burden, despite decades of targeted interventions. The World Health Organization estimates that in 2022 there were over 249 million malaria cases globally, resulting in more than 600,000 deaths, with the majority occurring in African countries \citep{who2023world}.
In Ghana, malaria is hyperendemic and transmission persists throughout the year, though it peaks during the rainy seasons. The disease accounts for a substantial proportion of hospital outpatient visits and inpatient admissions, particularly among children under five years of age and pregnant women \citep{GSS2019MIS}.
While the implementation of vector control measures, improved diagnostics, and treatment campaigns has contributed to a gradual reduction in prevalence, national surveillance data indicate marked inter-annual and regional variability in malaria admissions. These persistent fluctuations underscore the complex, nonlinear, and spatially heterogeneous nature of malaria dynamics in Ghana.

The reliable quantification and forecasting of malaria admissions are essential for informed planning of healthcare resources and preventive interventions.
However, the available datasets used for such assessments are typically limited in temporal resolution and spatial detail.
Most health-facility records and national surveillance data provide aggregated annual or monthly summaries, often spanning only a decade, as is the case for Ghana’s 2014--2023 malaria admission data. Such short or moderately sized time series constrain the applicability of classical statistical and machine-learning models. Linear and autoregressive frameworks (such as ARIMA\footnote{Autoregressive Integrated Moving Average.}) require assumptions of stationarity and consistent seasonal cycles, which are rarely satisfied in malaria datasets characterised by irregular peaks, long-tailed fluctuations, and structural breaks caused by climatic variability or intervention rollouts \citep{Lunde2013}.
On the other hand, data-intensive machine-learning models, though flexible, demand larger training samples to achieve stability and generalisability \citep{McNeish2016}.
In low-data regimes, they tend to overfit, yielding implausible extrapolations and inflated uncertainty.

Hospital-based malaria admission data also present intrinsic representativeness challenges. Admission counts reflect both epidemiological transmission and contextual factors such as access to healthcare, reporting completeness, and facility catchment dynamics. Consequently, raw admission trends can obscure true transmission intensity or exaggerate local fluctuations due to variations in service utilisation \citep{Alegana2020}.
Spatial delineation of health-facility catchments and adjustment for reporting biases have been shown to alter the perceived geography of malaria burden \citep{Donkor2021}.
Ignoring such spatial heterogeneity risks misinterpreting temporal patterns as epidemiological trends when they may instead reflect reporting artefacts.
A joint temporal–spatial approach that separates the temporal signal from the spatial variability, therefore becomes crucial to reveal underlying dynamics and to identify consistent high-burden or highly variable regions.

Malaria transmission dynamics are also shaped by multiple nonlinear processes. These include rainfall-driven mosquito breeding cycles, temperature-dependent parasite development, and socio-behavioural drivers such as preventive bed-net use or treatment uptake \citep{cibulskis2016malaria}.
Such interactions produce epidemic curves with abrupt inflexion points, plateaux, and intermittent surges that cannot be adequately captured by simple linear regressions or fixed-seasonal autoregressive models. Even where covariates are available, parametric forms may fail to express complex phase relationships among drivers. A modelling framework that allows flexible curvature, accommodates stochastic variability, and directly quantifies parameter uncertainty is therefore needed to provide credible inference under these conditions.

Bayesian inference provides a framework that combines prior information with observed data, allowing Bayesian models to effectively handle the uncertainty associated with small samples and reflect it within predictive intervals.
This approach is particularly valuable for health surveillance systems with limited temporal depth or incomplete spatial coverage \citep{yang2020bayesian,edition2013bayesian,carlin2008bayesian}. 
A Bayesian nonlinear structure, such as a hybrid cubic component augmented by a bounded nonlinear term, can capture smooth long-term changes while flexibly representing transient anomalies. Posterior sampling techniques, such as ensemble Markov chain Monte Carlo (MCMC), allow efficient exploration of parameter space and generation of full posterior distributions for key epidemiological quantities, rather than point estimates alone. This probabilistic perspective aligns with public-health needs for decision-making under uncertainty, where credible intervals provide more informative guidance than single deterministic forecasts.

To avoid ambiguity regarding scope, the modelling framework adopted in this study is deliberately parsimonious, reflecting the limited temporal depth of routine malaria surveillance data in Ghana. The emphasis is placed on Bayesian inference, uncertainty quantification, and epidemiological interpretability, rather than on maximising functional complexity. This design choice ensures stable estimation and credible probabilistic forecasting in a low-sample regime typical of national health-facility datasets.

In this study, we develop and apply a Bayesian nonlinear modelling framework to characterise malaria admission dynamics in Ghana over the period 2014--2023. 
The approach is designed to handle limited data while retaining interpretability and statistical robustness.
We model separate temporal trajectories for children under five years and individuals aged five years or older to capture demographic contrasts in burden and variability. Posterior inference is used to evaluate parameter uncertainty and to forecast near-future admissions for 2024--2026. 
The resulting posterior predictive distributions provide a measure of forecast reliability and enable probabilistic evaluation of potential increases or decreases in hospital admissions.

Beyond temporal inference, the study also examines spatial heterogeneity in malaria admission variability. Using district-level coefficients of variation and burden classes (high–high, high–low, low–high, and low–low), we identify geographical areas that deviate consistently from national trends. These diagnostics allow differentiation between districts characterised by high mean burden but stable variability and those with low mean burden but volatile admissions, providing operational insight into regions that may require strengthened surveillance or intervention focus.

The remainder of this paper is organised as follows. Section~\ref{sec:studyarea} describes the study area and data, outlining the spatial scope, demographic composition, and preprocessing of Ghana’s 2014--2023 malaria admission records. The analytical framework is then presented in Section~\ref{sec:method}, which covers the computation of district-level temporal variability and the formulation of the Bayesian hybrid cubic–damped oscillatory kernel model, estimated using an ensemble MCMC sampler. Section~\ref{sec:RNA} presents and discusses the empirical results, including analyses of spatio-temporal variability, posterior parameter estimates, model diagnostics, and probabilistic forecasts for 2024--2026, as well as their epidemiological and policy implications for malaria surveillance in Ghana. Finally, Section~\ref{sec:conc} summarises the principal insights, outlines methodological limitations, and proposes directions for future research and the integration of the framework within national health decision-support systems.

\section{Data and Study Area} \label{sec:studyarea}

\subsection{Study area} \label{sec:Sarea}

\begin{figure*}
\begin{minipage}[H]{0.85\linewidth}
\centering
\includegraphics[width=\textwidth]{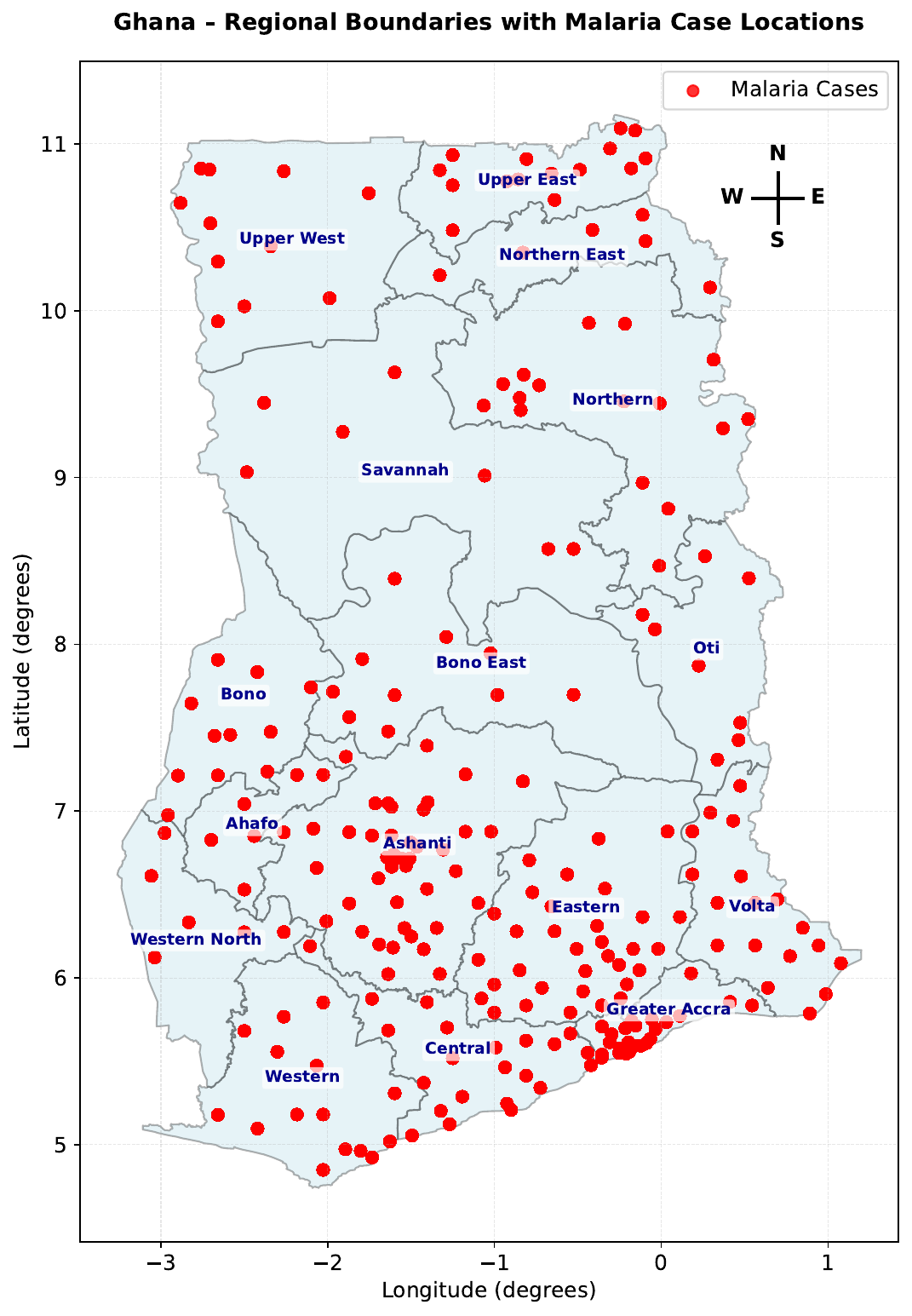} 
\end{minipage}
\caption{Administrative map of Ghana showing the sixteen regions and the spatial distribution of district-level malaria case-reporting locations used in this study. Red markers indicate district centroids associated with aggregated hospital admission records
}\label{fig:ghmap_f1}
\end{figure*}

Ghana is located along the Gulf of Guinea in West Africa, approximately between latitudes 4.5\textdegree N and 11.5\textdegree N and longitudes 3.5\textdegree W and 1.5\textdegree E.
It covers a total area of about $239,460\, \text{km}^2$, of which nearly $8,500 \, \text{km}^2$ are inland water bodies. Administratively, the country is divided into sixteen regions and $261$ districts, which together form the principal framework for governance and health-service delivery. 
This administrative structure, particularly the district level, provides the spatial unit for our analysis. Consequently, this study focuses on malaria case surveillance across all regions, with data aggregated at the district level to align with the operational level of health-service delivery. A visual overview of this framework is provided in Fig.~\ref{fig:ghmap_f1}, which presents a map of Ghana showing regional boundaries and the spatial distribution of malaria case-reporting sites.

The considerable north-south span of the country results in a diverse ecology that is a key determinant of malaria transmission heterogeneity. Ghana is typically grouped into three broad ecological belts: the Guinea Savannah in the north, the Forest zone in the centre, and the Coastal Savannah in the south \citep{adu2015spatiotemporal,de2010environmental}.
These zones exhibit distinct rainfall patterns and hydrological profiles, which directly shape the seasonality and intensity of malaria transmission. For instance, the northern regions generally experience a unimodal rainy season (May--October), while the middle and southern belts have bimodal rainfall peaks (April--June and September--November), influencing vector breeding and transmission cycles accordingly \citep{awine2018spatio}.

Against this backdrop of varied ecological and administrative landscapes, malaria remains endemic throughout Ghana, with a consistently high incidence rate reported among children under $10$ years of age, the most vulnerable demographic group \citep{adokiya2025perspectives,aidoo2024malaria,kolekang2022challenges}.
Despite ongoing control efforts, the disease exhibits pronounced spatial and seasonal variations, reflecting disparities in vector ecology, healthcare accessibility, and socioeconomic factors across regions. 
Urban areas, particularly Greater Accra, may report lower prevalence due to improved housing, drainage, and health infrastructure, whereas higher transmission persists in rural districts with limited access to preventive and curative services \citep{fobil2011neighborhood}.

\subsection{Data description}\label{sec:data}
The dataset used in this analysis comprises monthly malaria hospital admission records spanning 2014--2023, aggregated at the district level across all sixteen administrative regions of Ghana. Records were obtained from Ghana Health Service facility-based surveillance data compiled from routine malaria admissions nationwide.

Each record reports confirmed malaria admissions disaggregated into two paediatric age categories used in Ghana Health Service surveillance: children under five years ($<5$ years) and children aged five to nine years (5--9 years). Throughout this study, the notation \enquote{$\geq 5$ years} therefore refers exclusively to older children aged 5--9 years and does not include adults. These age groups represent epidemiologically distinct paediatric populations, reflecting the heightened vulnerability of children under five due to limited acquired immunity and the partial immunity that develops among older children \citep{awine2018spatio,Aregawi2017}. This stratification is well established in malaria surveillance across sub-Saharan Africa and avoids aggregation-induced masking of age-specific transmission dynamics \citep{who2023world,owusu2009epidemiology}.

To ensure analytical consistency across the study period, data cleaning procedures were undertaken to harmonise district names following administrative revisions (2018--2021), verify temporal coherence, and remove incomplete or duplicate entries. Monthly admission counts were aggregated into annual totals for temporal modelling, while retaining monthly resolution for spatio-temporal analyses. District centroid coordinates were appended to each record to support spatial visualisation.

\begin{figure*}
\begin{minipage}[H]{\linewidth}
\centering
\includegraphics[width=\linewidth]{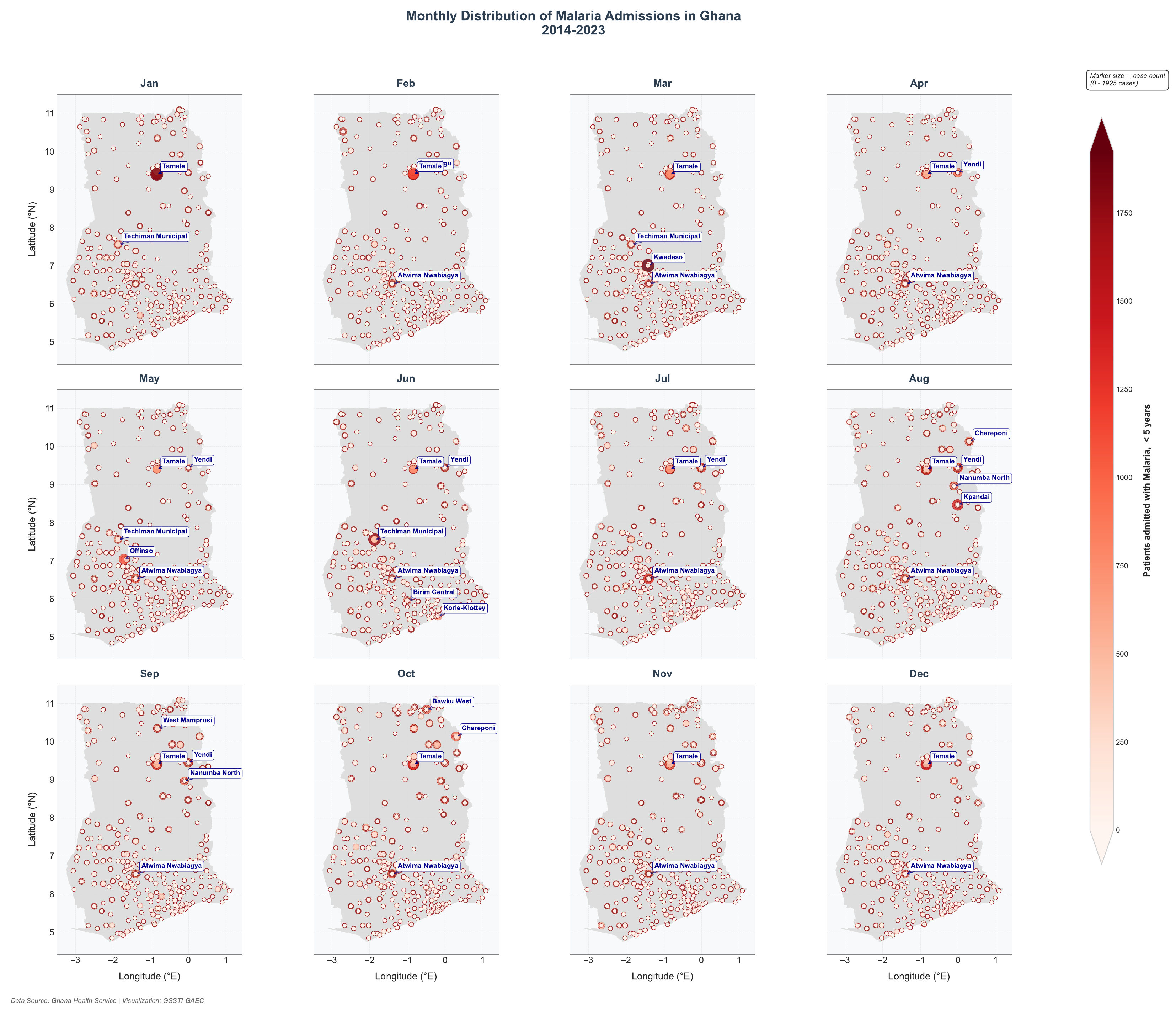} 
\end{minipage}
\caption{Monthly spatial distribution of malaria admissions among children under five years ($<$ 5) across Ghana (2014--2023). Each panel represents one calendar month, with marker size proportional to district-level admission counts. 
}\label{fig:monthly_malaria_hotspots_f2}
\end{figure*}

\begin{figure*}
\begin{minipage}[H]{\linewidth}
\centering
\includegraphics[width=\linewidth]{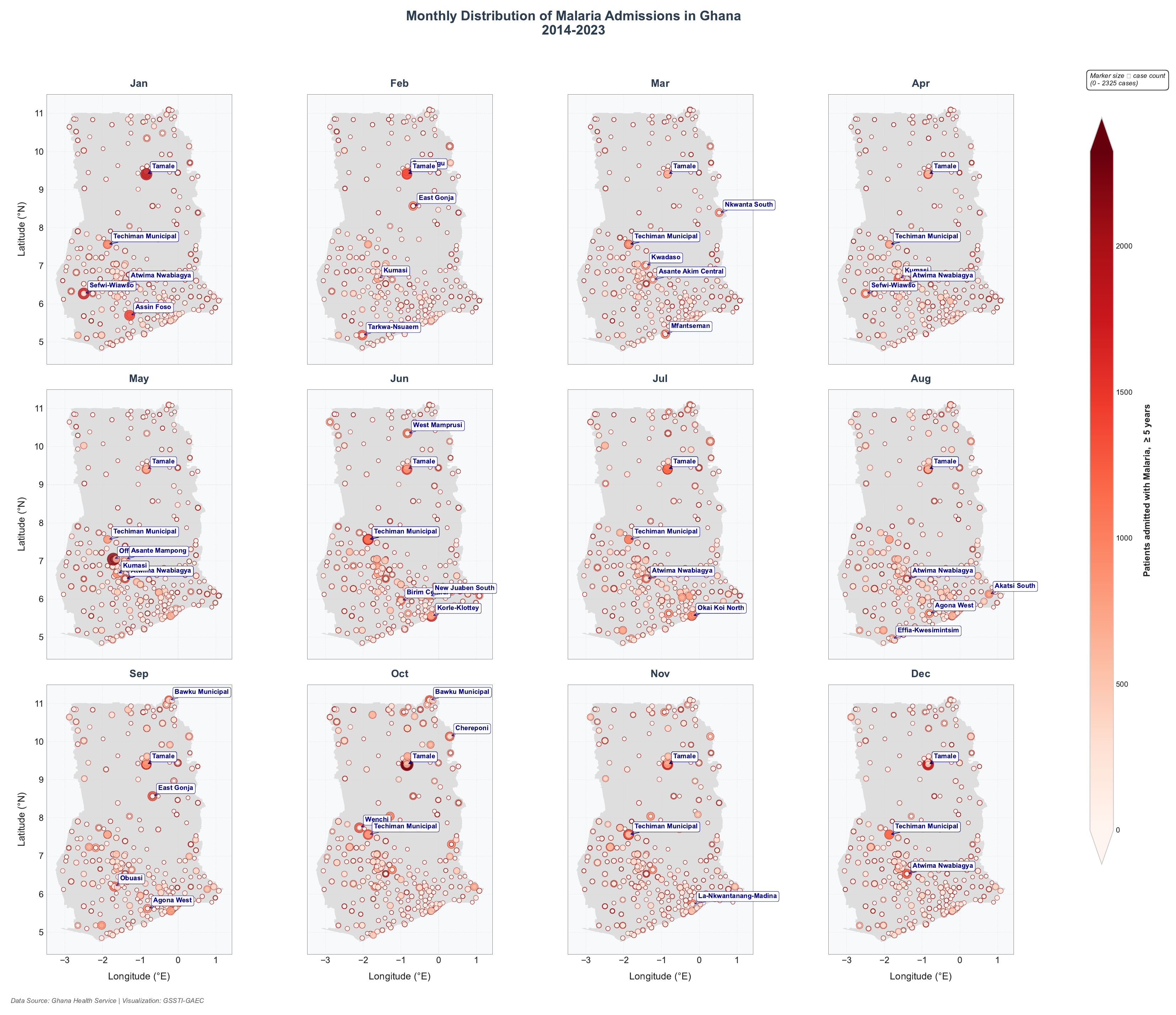} 
\end{minipage}
\caption{
Monthly spatial distribution of malaria admissions among individuals aged five years and above ($\geq 5$) across Ghana (2014--2023). Marker size reflects district-level admission counts for each calendar month.
}\label{fig:monthly_malaria_hotspots_Ov5}
\end{figure*}

Figures~\ref{fig:monthly_malaria_hotspots_f2} and~\ref{fig:monthly_malaria_hotspots_Ov5} present the spatial distribution of reported malaria admissions for the $<5$ years and $\geq 5$ years groups, respectively, disaggregated by calendar month from January to December. The maps provide a descriptive overview of the geographic and seasonal structure of malaria admissions across Ghana. Higher admission intensities are observed in northern districts, including Tamale, Savelugu, Yendi, and parts of the Upper East and Upper West regions, consistent with the unimodal rainfall regime and extended transmission season characteristic of the Guinea Savannah ecological zone \citep{adu2015spatiotemporal,asare2017assessing}. In contrast, southern and coastal districts, such as Greater Accra and parts of the Western and Central regions, exhibit comparatively lower case densities, reflecting shorter transmission periods and improved access to malaria prevention and treatment infrastructure \citep{aidoo2024malaria,kolekang2022challenges}.

Across both age groups, admissions generally increase between May and August with the onset of the rainy season and peak between September and October, while January to March record the lowest counts during the dry season when vector breeding is limited. Nevertheless, reported cases persist throughout the year, particularly in northern and transition-zone districts, indicating perennial transmission in some localities. High-density foci observed in metropolitan areas such as Tamale, Techiman, and Kumasi further reflect the influence of population mobility and urban health service utilisation on reported admissions.

  \begin{figure}
\begin{minipage}[H]{\linewidth}
\centering
\includegraphics[width=\textwidth]{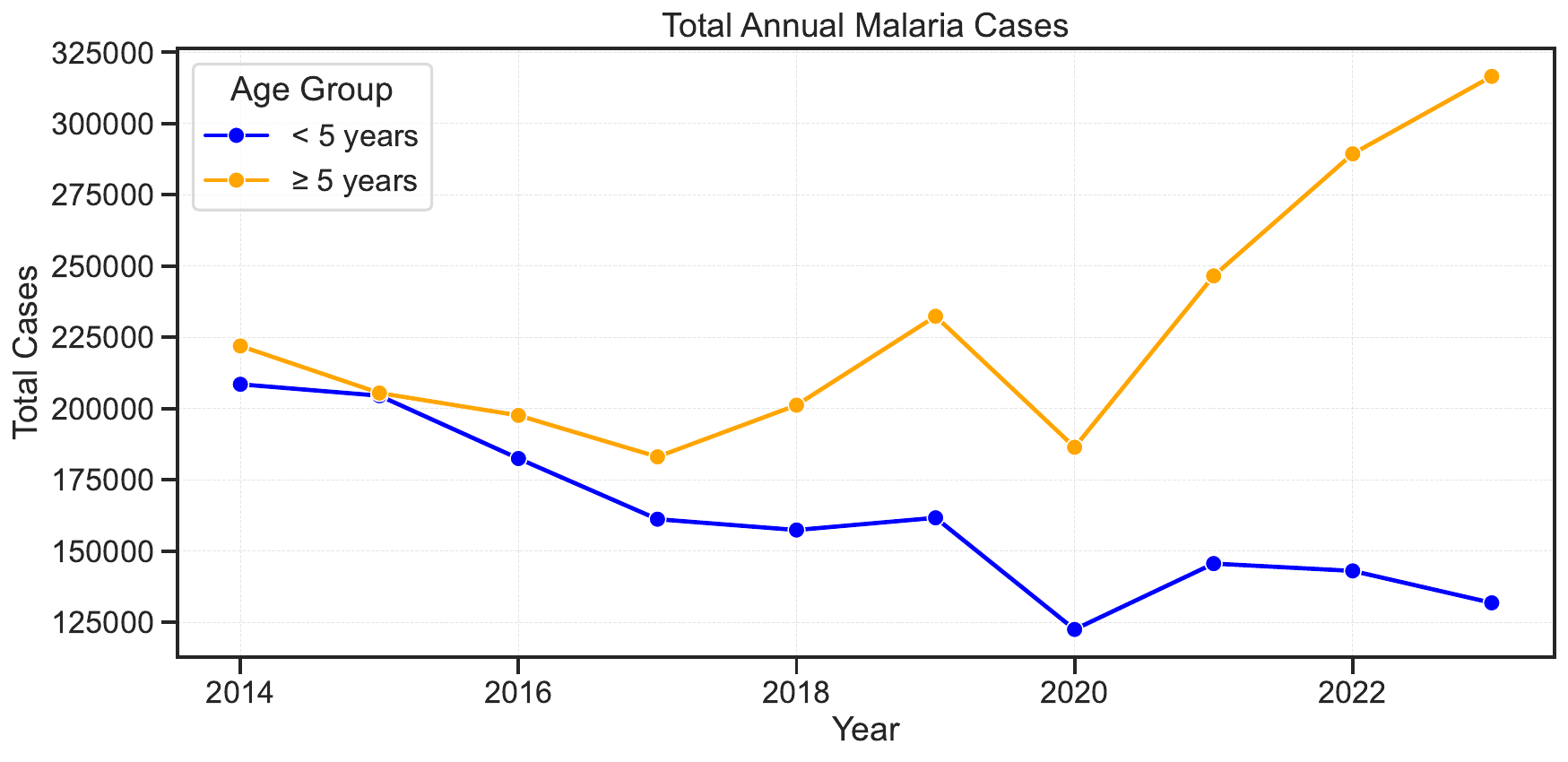} 
\end{minipage}
\caption{Total annual malaria admissions in Ghana from 2014--2023, disaggregated by age group ($<5$ years and $\geq5$ years). Lines represent aggregated national counts derived from monthly district-level records.
}\label{fig:ann_mal_cases}
\end{figure}

Figure~\ref{fig:ann_mal_cases} presents the annual total malaria case counts in Ghana from 2014 to 2023, disaggregated by age group ($<5$ years and $\geq 5$ years). The annual series, derived from monthly surveillance data, exhibit pronounced inter-annual variability, highlighting the non-linear and stochastic nature of malaria transmission dynamics. The $\geq 5$ years group displays higher-amplitude fluctuations, with notable peaks in 2019 and 2023, whereas the $<5$ years group shows a generally declining trend from 2014 to 2020 followed by a modest resurgence. The contrasting temporal behaviour between the two cohorts underscores the need for modelling frameworks capable of capturing age-specific heterogeneity and non-linear temporal responses. Overall, the observed non-linearity, age-specific heterogeneity, and pronounced inter-annual variability in the surveillance data, combined with the relatively short temporal span of the available records, necessitate a modelling framework that can explicitly represent parameter uncertainty and generate probabilistic predictions. A Bayesian approach is therefore adopted in subsequent analyses to accommodate overdispersed count data, quantify uncertainty in latent temporal dynamics, and provide credible predictive intervals suitable for inference and decision support.

\section{Methods} \label{sec:method}

\begin{figure}[H]
\centering
\begin{tikzpicture}[
    node distance=0.8cm,
    every node/.style={font=\small},
    box/.style={rectangle, draw, rounded corners, align=center, minimum width=3.8cm, minimum height=1.1cm},
    bigbox/.style={rectangle, draw, rounded corners, align=center, minimum width=5.2cm, minimum height=1.2cm},
    diamond/.style={diamond, draw, align=center, aspect=2, inner sep=1pt},
    arrow/.style={->, thick}
]

% -------------------- DATA --------------------
\node[box] (data) {%
\textbf{Observed Data}\\
District-level annual malaria admissions\\
(2014--2023), stratified by age group\\
($<5$, $\geq5$ years)
};

% -------------------- MODEL --------------------
\node[bigbox, below=of data] (model) {%
\textbf{Bayesian Model Specification}\\
Negative binomial likelihood for over-dispersed counts\\
Hybrid cubic--damped oscillatory temporal structure\\
Prior distributions on all parameters
};

% -------------------- INFERENCE --------------------
\node[bigbox, below=of model] (inference) {%
\textbf{Posterior Inference}\\
Affine-invariant ensemble MCMC sampling\\
Posterior $p(\boldsymbol{\theta}\mid \mathbf{y})$\\
Convergence diagnostics and trace inspection
};

% -------------------- PPC --------------------
\node[bigbox, below=of inference] (ppc) {%
\textbf{Posterior Predictive Analysis}\\
Generate replicated datasets\\
Posterior predictive checks (PPCs)\\
Uncertainty propagation
};

% -------------------- FORECAST --------------------
\node[bigbox, below=of ppc] (forecast) {%
\textbf{Forecasting}\\
Posterior predictive distributions\\
Short-term projections (2024--2026)\\
Widening uncertainty beyond observed period
};

% -------------------- DECISION SUPPORT --------------------
\node[bigbox, below=of forecast] (decision) {%
\textbf{Probabilistic Decision Support}\\
Uncertainty-aware risk interpretation\\
Age-specific burden comparison\\
Inform expert judgement under data constraints
};

% -------------------- ARROWS --------------------
\draw[arrow] (data) -- (model);
\draw[arrow] (model) -- (inference);
\draw[arrow] (inference) -- (ppc);
\draw[arrow] (ppc) -- (forecast);
\draw[arrow] (forecast) -- (decision);

\end{tikzpicture}
\caption{Schematic overview of the analytical workflow adopted in this study. The pipeline proceeds from preparation of district-level malaria admission data, through Bayesian model specification and posterior inference, to posterior predictive evaluation and uncertainty-aware forecasting. The final stage emphasises probabilistic decision support rather than automated rule-based optimisation.}
\label{fig:workflow}
\end{figure}
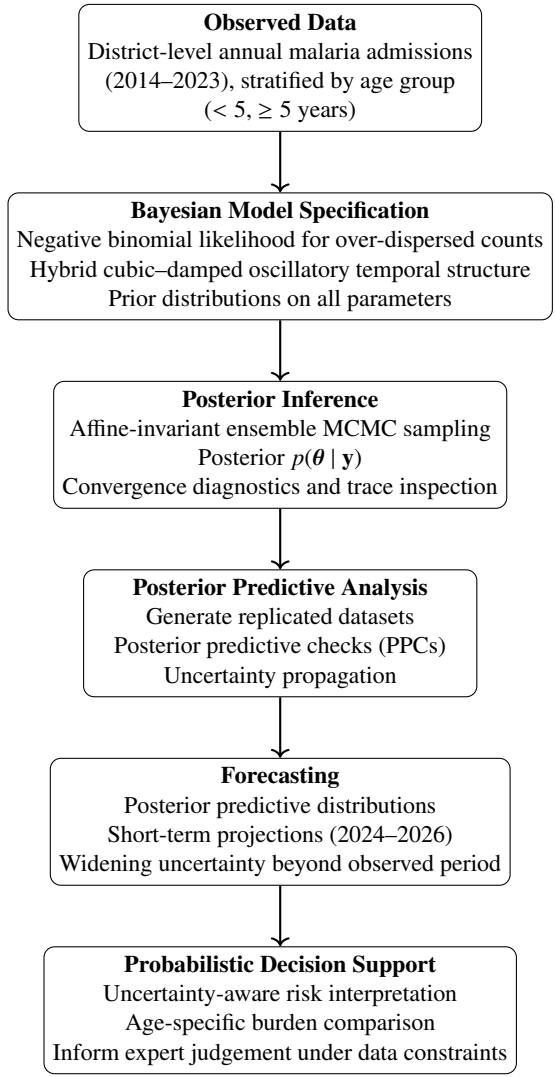

The complete analytical workflow adopted in this study is summarised in Fig.~\ref{fig:workflow}, illustrating the sequential progression from data preparation through Bayesian inference to uncertainty-aware forecasting and decision-support interpretation.

\subsection{Statistical assessment of spatial variability}
\label{sec:Cvs}

In national disease surveillance settings, administrative units often differ substantially in population size, healthcare access, and baseline disease burden. Under such conditions, absolute measures of dispersion alone (e.g., standard deviation) are inadequate for comparative analysis, because they scale with the mean and therefore confound variability with burden. To enable meaningful comparison of temporal variability across districts with heterogeneous malaria incidence levels, a scale-normalised measure is required \citep{bland2015introduction, abdi2010coefficient, van2005statistics}.

Let $y_{i,m}$ denote the number of reported malaria admissions in district $i$ during month $m$, with $m = 1, \ldots, M$. The expected monthly admission count for district $i$ is defined as
\begin{equation}
\bar{y}_i = \frac{1}{M}\sum_{m=1}^{M} y_{i,m},
\label{eq:mean_y}
\end{equation}
and the corresponding sample standard deviation is
\begin{equation}
s_i = \sqrt{\frac{1}{M-1}\sum_{m=1}^{M}(y_{i,m}-\bar{y}_i)^2}.
\label{eq:sample_std}
\end{equation}

\noindent
Following standard statistical practice, temporal variability was summarised using the coefficient of variation (CV) defined in \cref{eq:cv},
\begin{equation}
\mathrm{CV}_i = \frac{s_i}{\bar{y}_i},
\label{eq:cv}
\end{equation}
which expresses dispersion relative to the mean and yields a dimensionless index of relative variability. The CV facilitates direct comparison between districts with markedly different average malaria burdens and has been widely adopted in epidemiology and medical statistics for this purpose \citep{kirkwood2010essential}.

The coefficient of variation is a standard descriptive statistic and has been extensively used in public-health surveillance to characterise relative temporal instability, reporting consistency, and heterogeneity across spatial units. In the present application, elevated $\mathrm{CV}_i$ values identify districts where month-to-month malaria admissions exhibit large fluctuations relative to their mean level, whereas low $\mathrm{CV}_i$ values indicate districts with comparatively stable temporal admission profiles.

Importantly, this classification has direct implications for stochastic interpretation. Districts with high CV values are dominated by irregular, high-amplitude deviations from their mean behaviour, consistent with stronger stochastic influences arising from episodic outbreaks, seasonal forcing, reporting variability, or population mobility. In contrast, low-CV districts exhibit reduced relative variance and more regular temporal structure, indicating that observed admissions are more tightly concentrated around their expected level. Distinguishing between these regimes is necessary for interpreting subsequent probabilistic models, as it clarifies whether observed variability is primarily driven by stochastic fluctuations or by smoother, structurally persistent dynamics.

%% 3.2
\subsection{Bayesian modelling of temporal malaria admission dynamics}
\label{sec:bayesian_model}

Temporal malaria admissions exhibit pronounced nonlinear and nonstationary behaviour, shaped by intervention coverage, demography, climatic variability, and health-service utilisation across Ghana. In national surveillance settings characterised by short time series and heterogeneous reporting processes, conventional linear or high-capacity forecasting models are either overly restrictive or statistically under-identified. To address these constraints, a Bayesian modelling framework was developed to:
\begin{enumerate}[i.]
    \item estimate smooth yet flexible nonlinear temporal profiles for two age groups (\textless5 years and $\geq$5 years);
    \item quantify parameter uncertainty and posterior dependence structures;
    \item generate uncertainty-aware probabilistic forecasts for 2024--2026; and
    \item assess model adequacy using likelihood-based and posterior predictive diagnostics.
\end{enumerate}

\noindent
The Bayesian formulation is particularly suited to sparse health surveillance data, as it enables coherent uncertainty propagation and avoids overfitting through explicit prior regularisation \citep{carlin2008bayesian,gelman2013bayesian}.

Let $y_t$ denote the reported malaria admissions in year $t$. Because empirical variance substantially exceeded the mean, an over-dispersed \textit{negative-binomial} likelihood was adopted:
\begin{equation}
    y_t \sim \mathrm{NegBin}(\mu_t,\phi),
    \qquad
    \mathrm{Var}(y_t) = \mu_t + \frac{\mu_t^{2}}{\phi},
    \label{eq:nb}
\end{equation}

\noindent
where $\mu_t$ is the expected admission rate and $\phi > 0$ is the dispersion parameter controlling extra-Poisson variability. The negative binomial distribution is widely used in epidemiology and public-health surveillance to model count data exhibiting aggregation, unobserved heterogeneity, and reporting variability, conditions under which Poisson assumptions are violated 
\citep{hilbe2011negative, lloyd2005superspreading}. 
Larger values of $\phi$ imply weaker overdispersion and hence more stable variance around the mean. In the limit $\phi \to \infty$, \cref{eq:nb} reduces to the Poisson case, which is unrealistic for the present data given the observed temporal volatility and episodic surges in admissions.

Thus, the finite $\phi$ estimated in this study provides an interpretable measure of stochasticity and aggregation in reported malaria admissions, rather than attributing all deviations to deterministic structure. Uncertainty arising from reporting completeness, diagnostic practices, and health-facility catchment heterogeneity is not modelled explicitly, but is implicitly absorbed through overdispersion and posterior uncertainty quantification; a detailed discussion of these methodological scope considerations is provided in Section~\ref{subsec:scope}. Temporal indices were normalised to $x_t \in [-1,1]$ and counts scaled to $[0,1]$ to stabilise numerical optimisation and sampling.

The deterministic trend $\mu_t$ combines a low-order polynomial baseline with an additive damped oscillatory component:
\begin{equation}
\mu_t =
a x_t^{3} + b x_t^{2} + c x_t + d
+ A \tanh(\sigma z_t)\sin(\rho z_t)\exp(-\beta z_t^{2}),
\label{eq:mean}
\end{equation}
where $z_t = x_t - x_0$ recentres the process in time. The cubic polynomial $(a,b,c,d)$ captures smooth long-term evolution, while the additive kernel term
$K(z_t;\boldsymbol{\theta}_K) = A \tanh(\sigma z_t)\sin(\rho z_t)\exp(-\beta z_t^2)$
introduces bounded, non-periodic oscillations governed by amplitude $A$, scaling $\sigma$, frequency $\rho$, and decay rate $\beta$.

This hybrid structure is motivated by the need to balance flexibility and identifiability in short time series. Unlike seasonal autoregressive models, which impose fixed periodicity, or Gaussian processes and machine-learning models, which require substantially larger sample sizes to constrain hyperparameters, the proposed formulation allows transient deviations from long-term trends without enforcing strict cyclic behaviour or incurring overfitting \citep{hyndman2018forecasting, McNeish2016}. Similar damped or kernel-augmented formulations have been shown to capture irregular epidemiological and environmental dynamics under limited data conditions \citep{canals2025nonlinear,mielke2024}.

Weakly informative priors were assigned to all parameters to constrain implausible values while preserving data-driven learning:
\begin{equation}
\begin{aligned}
a,b,c,d &\sim \mathcal{N}(0,10), \\
A &\sim \mathcal{N}(0,5), \\
\sigma &\sim \mathrm{HalfNormal}(0,5), \\
\rho &\sim \mathrm{HalfCauchy}(0,10), \\
\beta &\sim \mathrm{HalfNormal}(0,5), \\
x_0 &\sim \mathcal{U}(\min(x),\max(x)), \\
\phi &\sim \mathrm{HalfCauchy}(0,5).
\end{aligned}
\label{eq:priors}
\end{equation}

\noindent
The joint posterior distribution follows directly from Bayes’ theorem:
\begin{equation}
p(\boldsymbol{\theta} \mid y)
\propto
\left[ \prod_t p(y_t \mid \mu_t(\boldsymbol{\theta}), \phi) \right]
p(\boldsymbol{\theta}),
\label{eq:posterior}
\end{equation}
where $\boldsymbol{\theta} = (a,b,c,d,A,\sigma,\rho,\beta,x_0,\phi)$. Posterior samples from \cref{eq:posterior} enable coherent uncertainty quantification for both parameter inference and predictive forecasting.

\subsubsection{Bayesian inference using the affine-invariant ensemble sampler}
\label{subsec:inference}

Posterior inference was conducted using the affine--invariant ensemble Markov Chain Monte Carlo (MCMC) sampler introduced by \citet{goodman2010ensemble} and implemented in the \texttt{emcee} library\footnote{\url{https://emcee.readthedocs.io/en/stable/}} \citep{foreman2013emcee}. 
This class of samplers is specifically designed for target distributions with strong parameter correlations and unknown or poorly scaled geometries, conditions that commonly arise in nonlinear hierarchical and mechanistic models. 
In contrast to conventional single-chain random-walk Metropolis--Hastings or Gibbs samplers, affine--invariant ensemble methods are insensitive to linear transformations of parameter space, leading to more robust and efficient exploration without extensive manual tuning \citep{goodman2010ensemble}.

The ensemble sampler evolves multiple interacting chains, referred to as \emph{walkers}, in parallel. 
Each walker proposes updates based on the relative positions of other walkers, allowing collective adaptation to the local geometry of the posterior distribution. 
This property is particularly advantageous for the present hybrid cubic--DOK model, whose parameters exhibit nontrivial posterior dependencies between trend, kernel, and dispersion components. 
Compared to Hamiltonian Monte Carlo (HMC) methods, which require differentiable likelihoods and careful tuning of step sizes and mass matrices, the affine--invariant ensemble sampler offers a simpler and more robust alternative for moderately high-dimensional nonlinear models with limited data \citep{betancourt2017conceptual}.

The ensemble MCMC framework provides a principled basis for uncertainty quantification in malaria admission dynamics. 
Rather than producing a single point estimate, the sampler yields full posterior distributions for all model parameters, from which credible intervals and posterior predictive envelopes can be derived. 
This probabilistic representation is essential for decision-support applications in public health, where planners must assess not only expected trends but also the range of plausible future outcomes under uncertainty \citep{gelman2013bayesian,carlin2008bayesian}.

Initial walker positions were centred on deterministic maximum-likelihood estimates obtained using \texttt{lmfit}\footnote{\url{https://lmfit.github.io/lmfit-py/}}, ensuring that the ensemble was initialised in a high-probability region of parameter space and reducing burn-in time. 
Seventy walkers ($N_{w}=70$) jointly explored the nine-dimensional parameter vector $\boldsymbol{\theta} = (a,b,c,d,A,\log\sigma,\rho,\beta,x_0)$ for $N_s = 2000$ iterations, with the first 300 samples discarded as burn-in. 
The choice of walker count exceeds the commonly recommended minimum of $2d$, where $d$ is the number of parameters, promoting stable ensemble behaviour and efficient mixing \citep{foreman2013emcee}. 
Mean acceptance fractions ranged between $0.28$ and $0.47$, consistent with efficient sampling, and the average integrated autocorrelation time was below 120 steps across parameters, indicating satisfactory convergence and effective exploration of the posterior.

The overall logic of the affine--invariant ensemble sampler, as applied in this study, is summarised schematically in Algorithm~\ref{alg:emcee}. 
Each walker proposes a new parameter vector by drawing a scaled ``stretch'' from another walker's position, evaluates the likelihood ratio of the proposed versus current state, and accepts the move according to the Metropolis criterion \citep{metropolis1953,hastings1970}. 
Through repeated iterations, the ensemble collectively approximates the joint posterior distribution $p(\boldsymbol{\theta}\mid y)$, from which uncertainty-aware inferences on malaria admission trajectories are obtained.

\begin{algorithm}[t!]
\caption{Affine-invariant ensemble MCMC for parameter estimation}
\label{alg:emcee}
\SetKwInOut{Input}{Input}
\SetKwInOut{Output}{Output}
\Input{Normalised data $y_t$; model $\mu_t(\boldsymbol{\theta})$; log-likelihood $\mathcal{L}(\boldsymbol{\theta})$; number of walkers $N_w$; steps $N_s$}
\Output{Posterior samples $\{\boldsymbol{\theta}_i^{(k)}\}$}

Initialise $N_w$ walkers around maximum-likelihood estimate $\boldsymbol{\theta}_0$\;
\For{$k \leftarrow 1$ \KwTo $N_s$}{
  \ForEach{walker $i$}{
    Select another walker $j \neq i$ uniformly at random\;
    Draw stretch factor $z \sim g(z)=1/\sqrt{z}$ on $[1/a,a]$\;
    Propose $\boldsymbol{\theta}' = \boldsymbol{\theta}_j + z(\boldsymbol{\theta}_i - \boldsymbol{\theta}_j)$\;
    Compute acceptance probability:
    \[
      q = \min\left(1, z^{d-1}
      \frac{\mathcal{L}(\boldsymbol{\theta}')\,p(\boldsymbol{\theta}')}{\mathcal{L}(\boldsymbol{\theta}_i)\,p(\boldsymbol{\theta}_i)}\right);
    \]
    Accept $\boldsymbol{\theta}'$ with probability $q$, otherwise retain $\boldsymbol{\theta}_i$.
  }
}
Discard burn-in samples and flatten remaining chains to obtain $\{\boldsymbol{\theta}_i^{(k)}\}$.
\end{algorithm}

Convergence was assessed using multiple complementary diagnostics. 
Visual inspection of trace plots and marginal posterior densities was used to detect non-stationarity or multimodality, while quantitative diagnostics included Gelman--Rubin statistics ($\hat{R} < 1.05$) and effective sample sizes exceeding 500 per parameter, consistent with recommended best practices for Bayesian computation \citep{gelman2013bayesian,vehtari2021rank}.
Posterior summaries (means, standard deviations, and $95\%$ credible intervals) were extracted for all parameters.
%%

% ==========================================================
\subsubsection{Model evaluation metrics}
\label{subsec:model_eval}

Model adequacy was evaluated using a set of complementary goodness-of-fit, parsimony, and predictive accuracy metrics computed from the posterior median predictions $\hat{y}_t$. In a Bayesian setting with limited temporal depth, no single statistic is sufficient to characterise model performance; accordingly, multiple diagnostics were employed to assess calibration, explanatory power, and relative model adequacy in a transparent and reproducible manner \citep{gelman2013bayesian,vehtari2021rank}.

The chi-square and reduced chi-square statistics were used as global measures of discrepancy between observed and fitted values:
\begin{equation}
\chi^{2} = \sum_{t} \frac{(y_t - \hat{y}_t)^{2}}{\hat{y}_t}, 
\qquad
\chi_{\nu}^{2} = \frac{\chi^{2}}{n - p},
\label{eq:chisq}
\end{equation}
where $n$ is the number of observations and $p$ is the number of free parameters. While originally developed in frequentist contexts, chi-square-based measures remain useful descriptive diagnostics for assessing overall lack-of-fit in count data models when interpreted cautiously and in conjunction with posterior uncertainty \citep{cameron2013regression}.

Model parsimony was assessed using the Akaike Information Criterion (AIC) and the Bayesian Information Criterion (BIC):
\begin{equation}
\mathrm{AIC} = 2p - 2\ln(\hat{L}), 
\qquad 
\mathrm{BIC} = p\ln(n) - 2\ln(\hat{L}),
\label{eq:aicbic}
\end{equation}
where $\hat{L}$ denotes the maximised likelihood. Although AIC and BIC are not fully Bayesian quantities, they are widely used as approximate criteria for comparing relative model adequacy and complexity, particularly when competing low-dimensional models are nested or closely related \citep{konishi2008information, burnham2002model}. In the present study, these criteria are employed as descriptive tools to assess whether the added flexibility of the hybrid cubic--DOK structure is justified relative to simpler alternatives.

Overall predictive accuracy was summarised using the coefficient of determination ($R^2$) and the root-mean-square error (RMSE):
\begin{equation}
R^{2} = 1 - \frac{\sum_t (y_t - \hat{y}_t)^2}
{\sum_t (y_t - \bar{y})^2},
\qquad
\mathrm{RMSE} = \sqrt{\frac{1}{n} \sum_t (y_t - \hat{y}_t)^2}.
\label{eq:fit}
\end{equation}
These metrics provide intuitive summaries of explanatory power and average deviation between observed and fitted values. In line with Bayesian best practice, they are reported as descriptive measures of in-sample adequacy rather than as indicators of out-of-sample generalisation, which is constrained by the limited temporal extent of the data \citep{gelman2013bayesian,vehtari2021rank}.

Collectively, this suite of evaluation metrics supports a balanced assessment of model performance, emphasising interpretability, uncertainty awareness, and robustness under data-limited surveillance conditions. This approach aligns with the study’s objective of providing probabilistic decision-support information rather than optimising predictive accuracy in a purely data-driven forecasting setting.

\subsubsection{Methodological scope and limitations}
\label{subsec:scope}

The Bayesian hybrid cubic–damped oscillatory framework proposed in this study is intentionally low-dimensional, reflecting the limited temporal depth of routine malaria admission data in Ghana. With only ten annual observations per age cohort, highly flexible or high-capacity time-series models such as seasonal autoregressive processes, Gaussian processes, or data-intensive machine-learning approaches are statistically under-identified and prone to overfitting or unstable extrapolation when applied to national surveillance data. The modelling strategy adopted here therefore prioritises parameter identifiability, inferential stability, and coherent uncertainty quantification over functional complexity. Importantly, simpler formulations, such as a purely cubic trend model, are nested within the proposed structure as limiting cases when the oscillatory amplitude is negligible.

The framework should not be interpreted as deterministic curve fitting. Malaria admissions are modelled as over-dispersed count data via a negative binomial likelihood, allowing stochastic variability and aggregation effects to be explicitly inferred rather than absorbed into residual noise. The damped oscillatory kernel introduces bounded, non-periodic deviations from the long-term cubic trend, enabling transient fluctuations to be captured without imposing rigid seasonal or cyclic assumptions. This structure is particularly suited to malaria surveillance data, where irregular intervention effects, climatic anomalies, and reporting heterogeneity can induce short-lived departures from otherwise smooth trends. However, the present analysis does not explicitly model reporting completeness, diagnostic misclassification, or health-facility catchment changes, as consistent auxiliary metadata quantifying these processes are not available for the full study period; inference is therefore conditional on reported admissions as observed. The framework is intended for probabilistic risk inference using routinely available surveillance data, rather than for mechanistic attribution of specific transmission drivers.

Formal out-of-sample validation is constrained by the limited temporal depth of annual surveillance data; accordingly, model adequacy is assessed using posterior predictive checks and uncertainty propagation rather than conventional train–test evaluation. As a result, in-sample fit metrics are reported as descriptive summaries of model adequacy rather than as evidence of generalisation performance.

The proposed framework is designed as a statistically defensible, uncertainty-aware probabilistic decision-support system for public-health risk assessment, rather than as a classical rule-based expert system, a learning-based AI sequence model, or a mechanistic transmission or high-resolution spatio-temporal model. In this study, \enquote*{decision support} refers to the provision of uncertainty-aware probabilistic forecasts intended to inform expert judgement under data-limited surveillance conditions, rather than the implementation of explicit allocation rules, trigger thresholds, or automated decision engines.

\section{Results and Discussion}\label{sec:RNA}
% % %%

The Results section is organised to mirror the analytical workflow outlined in Fig.~\ref{fig:workflow}. We first summarise spatial variability metrics derived directly from the observed data. We then present posterior inference results for the Bayesian temporal model, followed by posterior predictive evaluation and short-term forecasting. Finally, we interpret age-specific patterns and uncertainty in the context of probabilistic decision support.

\subsection{Observed spatial variability in malaria admissions}\label{sec:heatmaps}

\begin{figure*}
\begin{minipage}[H]{\linewidth}
\centering
\includegraphics[width=\textwidth]{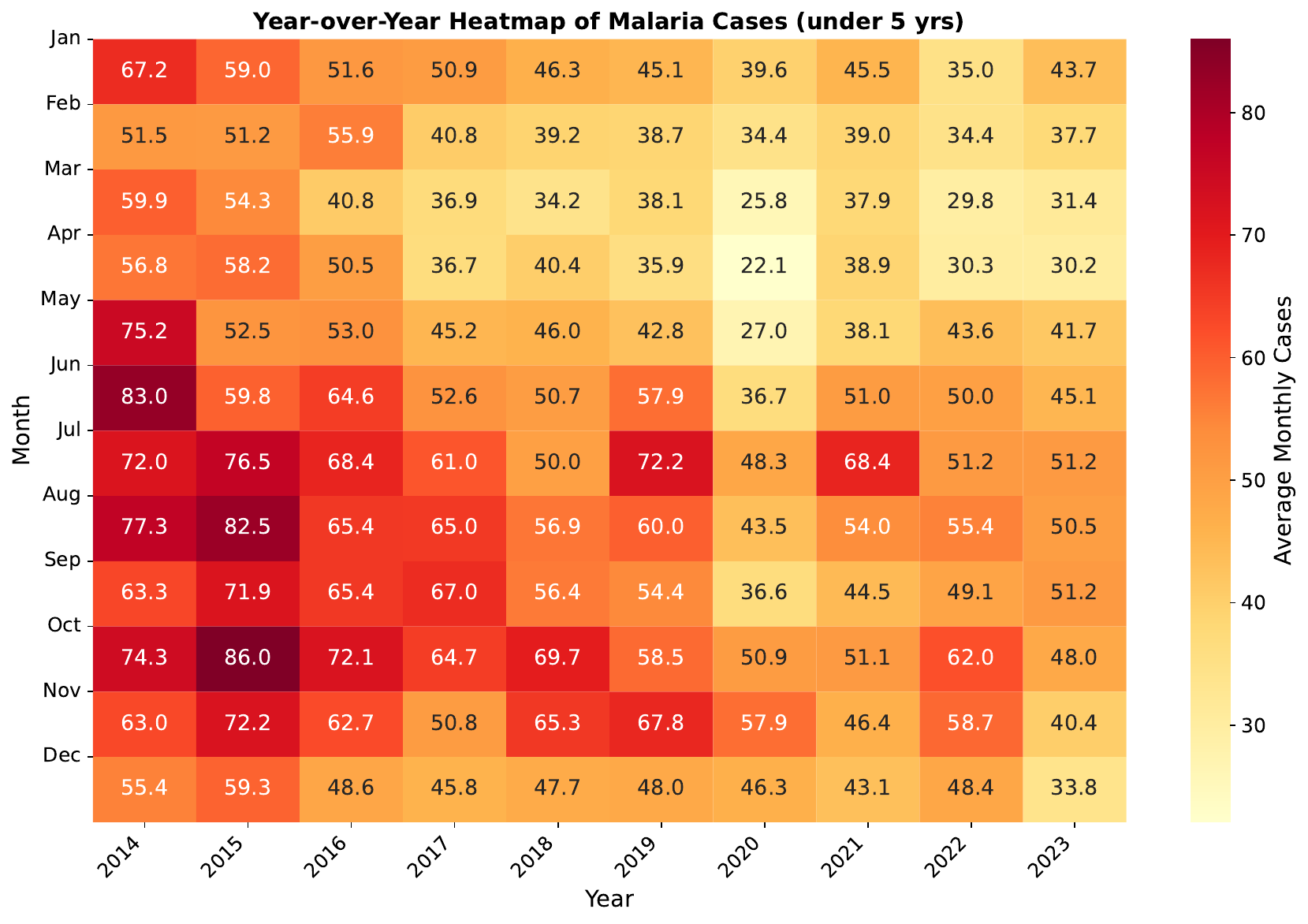} 
\end{minipage}
\caption{This heatmap shows a recurring seasonal pattern of malaria in young children ($< 5$ years) in Ghana, with admissions peaking during the rainy season (June-October). Crucially, while this seasonal cycle persists, the overall intensity has gradually decreased over the decade, indicating a positive decline in cases for this most vulnerable group.}\label{fig:malaria_heatmap_under5}
\end{figure*}

Figure~\ref{fig:malaria_heatmap_under5} illustrates the spatio-temporal distribution of malaria admissions among children ($<5$ years) across Ghana.
The matrix shows a strong and consistent seasonal pattern, with markedly higher mean monthly admissions from June to October, the main rainy period, and minima between December and March.
District average monthly values range from $22.1$ to $86.0$, with the January average declining from $67.2$ in 2014 to $43.7$ in 2023, reflecting an approximate $35\%$ reduction over the decade. These intra-annual oscillations align with rainfall-driven transmission cycles, where increases in vector breeding following the rains lead to case peaks after a short lag \citep{asare2017assessing,adu2015spatiotemporal}.
The gradual dampening of early-year intensities suggests that improved preventive coverage, including long-lasting insecticidal nets and seasonal chemoprevention, may be reducing the burden among younger children. Overall, the < 5 years heatmap depicts a temporally stable yet progressively moderating pattern of malaria admissions that mirrors the ecological and intervention context of Ghana’s endemic zones.

\begin{figure*}
\begin{minipage}[H]{\linewidth}
\centering
\includegraphics[width=\textwidth]{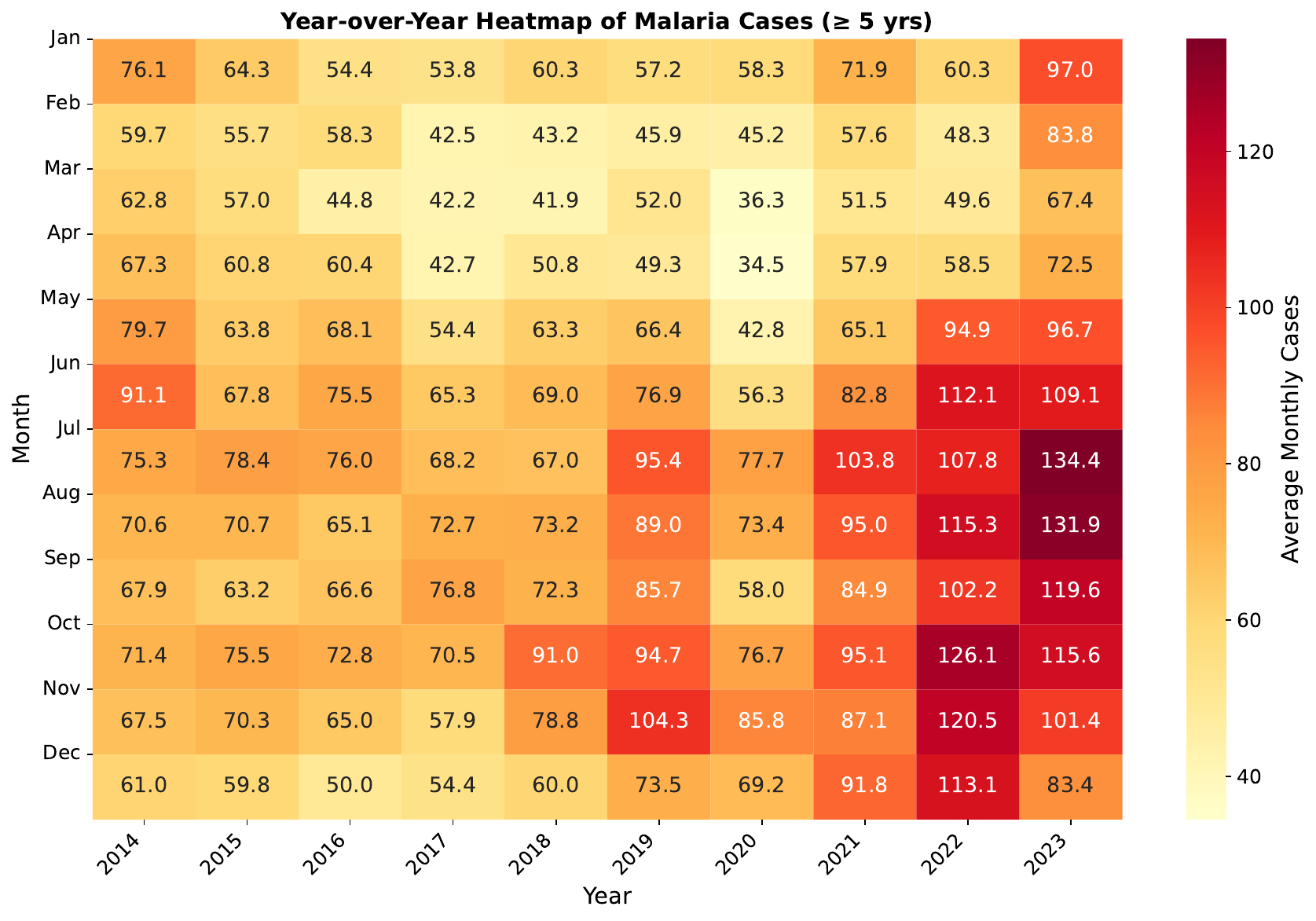} 
\end{minipage}
\caption{In contrast to younger children, this heatmap reveals that malaria in older children ($\geq 5$ years) is not only more intense but also more variable. Recent years show severe peaks, signalling a shift in the disease burden towards this older age group and highlighting persistent, climate-driven transmission cycles.}\label{fig:malaria_heatmap_over5}
\end{figure*}

The corresponding heatmap (in Fig.~\ref{fig:malaria_heatmap_over5}) for children $\geq$ 5 years shows substantially higher magnitudes and more pronounced inter-annual variability.
Average monthly district admissions span $34.5$--$134.4$, with peaks frequently exceeding $100$ cases between 2018 and 2023, particularly during the late-rainy-season months. The matrix maximum of $134.4$ represents roughly a $55 \%$ increase over the highest value in the $< 5$ years group, indicating a larger absolute disease burden in the older population. Transient declines observed around 2015 and 2020 are followed by sharp rebounds, coinciding with periods of intense rainfall and diminished vector-control efficacy reported across northern Ghana \citep{oheneba2022estimating,alout2017malaria}. The persistence of broad, high-intensity cells across successive months confirms sustained seasonal transmission rather than isolated outbreaks. These patterns, supported by emerging evidence of demographic shifts in malaria epidemiology \citep{savi2022modeling,owusu2009epidemiology}, suggest that while transmission remains strongly seasonal, the burden has increasingly extended into older age groups, highlighting the need for age-specific surveillance and temporally adaptive intervention strategies.

% \subsection{Spatial variability in malaria admission dynamics}\label{sec:cvar}

\begin{figure*}
\begin{minipage}[H]{0.85\linewidth}
\centering
\includegraphics[width=\textwidth]{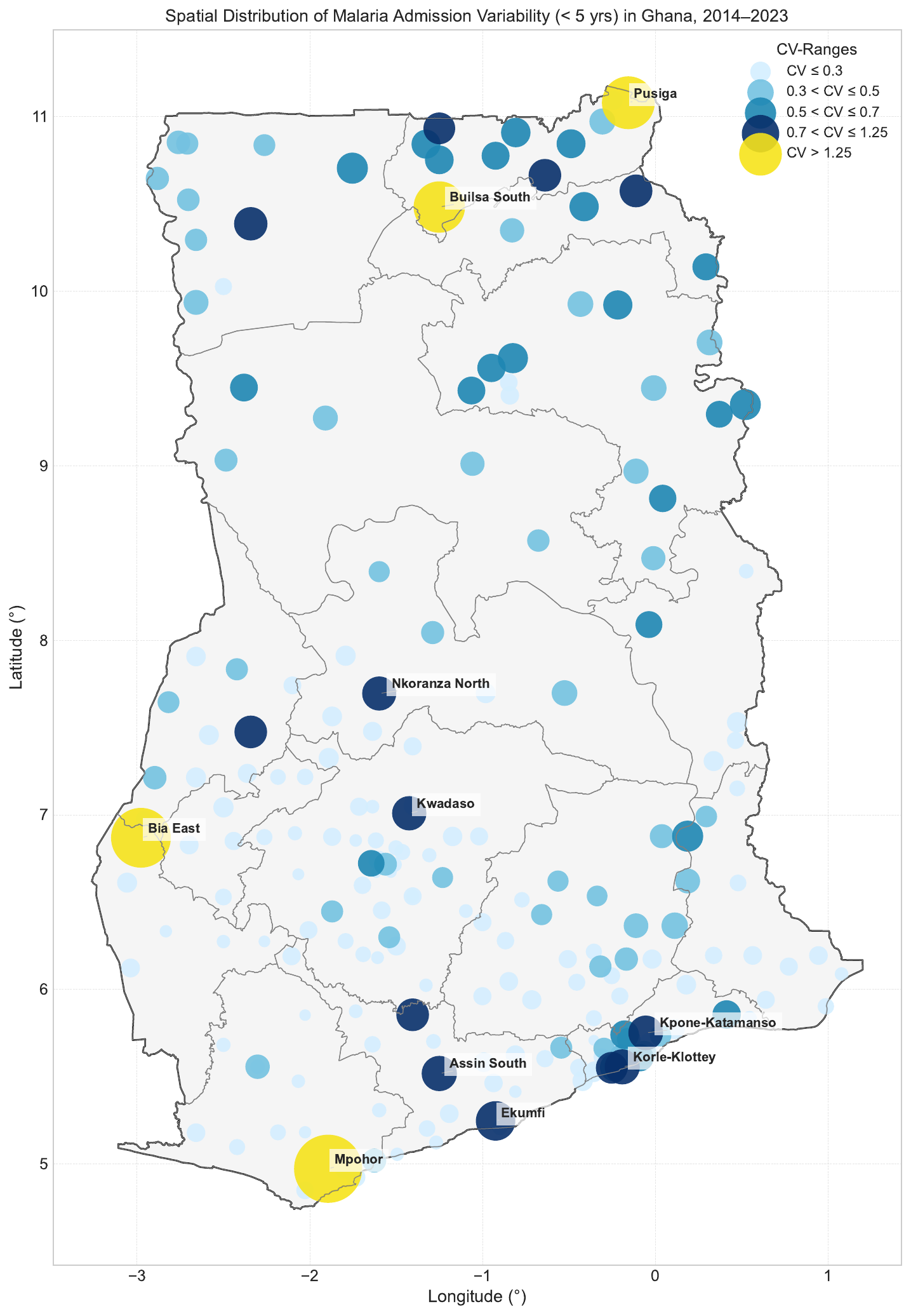} 
\end{minipage}
\caption{Spatial distribution of the coefficient of variation (CV) in monthly malaria admissions among children under five years ($<5$ years) across Ghana, 2014–2023.
District-level markers are scaled by mean monthly admissions and coloured by CV class ($\leq 0.3$, $0.3$--$0.5$, $0.5$--$0.7$, $0.7$--$1.25$, $> 1.25$). The map reveals pronounced spatial heterogeneity, with the highest temporal instability concentrated in smaller western and northern districts such as Mpohor and Bia East, while metropolitan areas including Kumasi and Accra exhibit markedly stable admission patterns.
}\label{fig:CV_malaria_variability_GH_yellow_highCV_u5}
\end{figure*}

Spatial patterns in the temporal variability of malaria admissions reveal marked heterogeneity across Ghana’s districts between 2014 and 2023.
The CV, which quantifies month-to-month instability in case numbers, varied by more than an order of magnitude from $< 0.07$ in the most stable districts to $> 3.3$ in the most volatile.
Figure~\ref{fig:CV_malaria_variability_GH_yellow_highCV_u5} depicts this spatial distribution for children under five years, while Figure 6 presents the corresponding pattern for individuals aged five years and above.
The detailed district-level statistics for both groups are summarised in Table \ref{tab:top10_cv} (top ten highest CVs) and Table \ref{tab:bottom10_cv} (ten lowest CVs).
Among children $< 5$ years, several districts stand out for their exceptionally high variability (CV $> 1.8$), notably Mpohor (CV $= 3.32$), Bia East ($2.55$), Pusiga ($1.99$), and Builsa South ($1.88$). 
These same districts report comparatively small mean monthly admissions, ranging from $\approx 0.5$ cases in Mpohor to $\approx 19$ cases in Pusiga, illustrating a high–low dynamic characterised by sporadic surges amid generally low baseline activity. 
Such erratic behaviour may reflect localised outbreak conditions, inconsistent data capture, or health-facility access disparities rather than sustained transmission intensity. 
In contrast, urban and peri-urban settings such as Kwadaso (CV $= 0.84$), Korle-Klottey ($0.88$), and Kpone-Katamanso ($0.86$) maintain consistently high mean monthly admissions ($\approx 70$--$630$ cases) with moderate variability, typifying high–high conditions in which case burdens remain persistently elevated but seasonally predictable.
The southern coastal and forest belts represented by districts including Kumasi, Ga West, and Wassa Amenfi East (CV $\leq 0.10$; Table \ref{tab:bottom10_cv}), exhibit remarkable temporal stability, suggesting regular transmission cycles and well-functioning reporting systems. Overall, the under-five pattern demonstrates that the volatility of malaria admissions is most pronounced in smaller, peripheral districts, while major metropolitan centres exhibit resilient, low-variance patterns that align with established service coverage and surveillance consistency.

\begin{figure*}
\begin{minipage}[H]{0.85\linewidth}
\centering
\includegraphics[width=\textwidth]{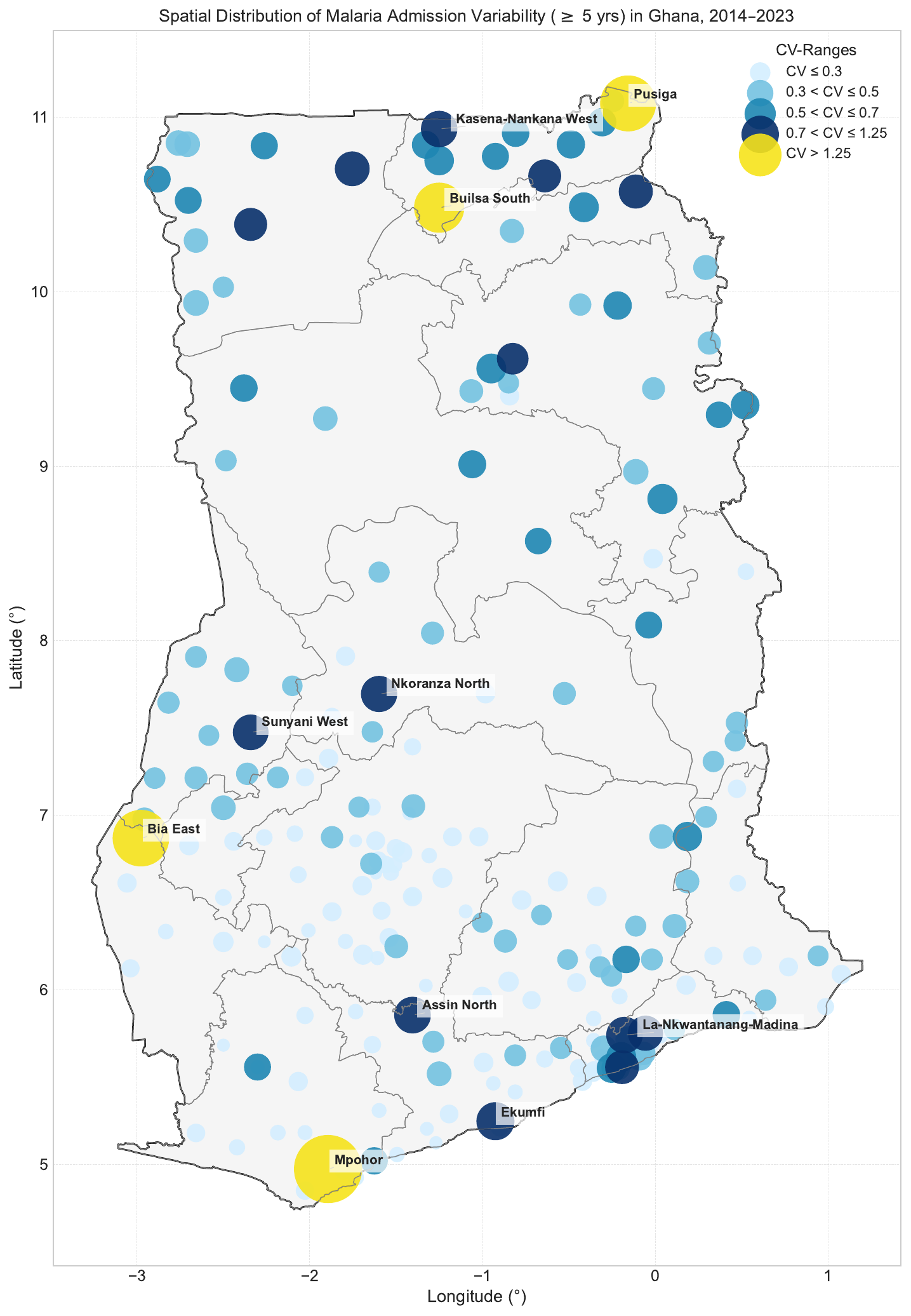} 
\end{minipage}
\caption{Spatial distribution of the coefficient of variation (CV) in monthly malaria admissions among individuals aged five years and above ($\geq 5$ years) across Ghana, 2014--2023.
Patterns mirror those observed in younger children, with persistent high-CV zones in Mpohor, Bia East, and Pusiga, contrasted by low-CV urban centres such as Kumasi, Bibiani-Anhwiaso-Bekwai, and Kwadaso. The figure highlights the spatial consistency of stable versus unstable districts across age groups and underscores the stronger temporal regularity within highly urbanised areas.
}\label{fig:CV_malaria_variability_GH_yellow_highCV_ovr5}
\end{figure*}

The pattern for individuals aged five years and above (Fig.~\ref{fig:CV_malaria_variability_GH_yellow_highCV_ovr5}) shows broadly similar spatial gradients but higher mean admissions and a modestly reduced CV spread. Again, Mpohor (CV $= 3.32$), Bia East ($2.28$), and Pusiga ($2.24$) emerge as the most unstable districts, mirroring their under-five behaviour and confirming persistent temporal irregularity across cohorts.
Average monthly admissions in these districts remain moderate ($\approx 6$--$58$ cases), reinforcing their classification as high–low variability zones.
In contrast, large urban and administrative centres such as La-Nkwantanang-Madina (CV $= 0.96$; mean $\approx 294$ cases), Kasena-Nankana West ($0.95$; $\approx 166$ cases), and Assin North ($0.97$; $\approx 100$ cases) display high–high profiles, combining substantial mean burdens with consistent temporal structure. 
The most stable districts for this age group (Table \ref{tab:bottom10_cv}) include Kumasi (CV $= 0.11$; mean $\approx 4771$ cases), Bibiani-Anhwiaso-Bekwai ($0.12$; $\approx 2620$ cases), and Atwima Nwabiagya North ($0.12$; $\approx 1871$ cases), all of which exhibit steady monthly admissions reflective of mature health systems, high diagnostic throughput, and regular reporting practices. 
Taken together, Figs.~\ref{fig:CV_malaria_variability_GH_yellow_highCV_u5}  and~\ref{fig:CV_malaria_variability_GH_yellow_highCV_ovr5} underscore a persistent spatial stratification in malaria admission variability: volatility and potential under-reporting dominate peripheral rural districts, while stable, predictable patterns prevail in urbanised areas with extensive healthcare infrastructure. The repetition of Mpohor, Bia East, and Pusiga among the most variable districts in both age strata highlights their recurrent instability and the need for strengthened surveillance and data-quality assurance in these localities.

\begin{table*}[htbp]
\centering
\caption{Top ten districts with the highest coefficients of variation (CV) in monthly malaria admissions for children under five years ($< 5$ years) and for individuals aged five years and above ($\geq 5$ years) across Ghana, 2014–2023.  The CV is defined as the ratio of the standard deviation to the mean of monthly admissions for each district.  Higher CV values denote greater temporal instability in malaria admissions.}
\label{tab:top10_cv}
\resizebox{\textwidth}{!}{%
\begin{tabular}{lccc @{\hspace{1.5cm}} lccc}
\hline
\multicolumn{4}{c}{\textbf{Children $< 5$ years}} & \multicolumn{4}{c}{\textbf{Children $\geq 5$ years}} \\
\hline
\textbf{District} & \textbf{Mean} & \textbf{Std} & \textbf{CV} & 
\textbf{District} & \textbf{Mean} & \textbf{Std} & \textbf{CV} \\
\hline
Mpohor                 & 0.50  & 1.66  & 3.32 &  Mpohor                 & 5.67  & 18.79  & 3.32 \\
Bia East               & 17.25 & 43.99 & 2.55 &  Bia East               & 8.92  & 20.37  & 2.28 \\
Pusiga                 & 19.42 & 38.67 & 1.99 &  Pusiga                 & 57.92 & 129.54 & 2.24 \\
Builsa South           & 13.75 & 25.88 & 1.88 &  Builsa South           & 15.42 & 27.73  & 1.80 \\
Ekumfi                 & 8.33  & 9.27  & 1.11 &  Ekumfi                 & 14.92 & 15.42  & 1.03 \\
Assin South            & 10.42 & 9.26  & 0.89 &  Assin North            & 99.58 & 96.13  & 0.97 \\
Korle-Klottey          & 293.92& 258.89& 0.88 &  La-Nkwantanang-Madina  & 294.00& 280.90 & 0.96 \\
Kpone-Katamanso        & 72.58 & 62.41 & 0.86 &  Kasena-Nankana West    & 166.25& 158.09 & 0.95 \\
Kwadaso                & 632.75& 532.65& 0.84 &  Nkoranza North         & 96.58 & 90.85  & 0.94 \\
Nkoranza North         & 40.92 & 34.03 & 0.83 &  Sunyani West           & 77.33 & 70.63  & 0.91 \\
\hline
\end{tabular}%
}
\end{table*}

\begin{table*}[htbp]
\centering
\caption{Top ten districts with the lowest coefficients of variation (CV) in monthly malaria admissions for children under five years ($< 5$ years) and for individuals aged five years and above ($\geq 5$ years) across Ghana, 2014–2023.  Lower CV values denote greater temporal stability in malaria admissions.}
\label{tab:bottom10_cv}
\resizebox{\textwidth}{!}{%
\begin{tabular}{lccc @{\hspace{1.5cm}} lccc}
\hline
\multicolumn{4}{c}{\textbf{Children $< 5$ years}} & \multicolumn{4}{c}{\textbf{Children $\geq 5$ years}} \\
\hline
\textbf{District} & \textbf{Mean} & \textbf{Std} & \textbf{CV} &
\textbf{District} & \textbf{Mean} & \textbf{Std} & \textbf{CV} \\
\hline
Kumasi                     & 1568.08 & 109.60 & 0.07 &  Kumasi                     & 4771.42 & 504.90 & 0.11 \\
Ga West                    & 221.25  & 18.50  & 0.08 &  Bibiani-Anhwiaso-Bekwai    & 2619.58 & 309.74 & 0.12 \\
Wassa Amenfi East           & 1169.25 & 111.63 & 0.10 &  Atwima Nwabiagya North     & 1871.17 & 227.14 & 0.12 \\
Oforikrom                  & 294.67  & 28.15  & 0.10 &  Wassa Amenfi West           & 1042.83 & 127.33 & 0.12 \\
Ledzokuku                  & 299.33  & 29.18  & 0.10 &  Kwadaso                    & 1500.50 & 191.46 & 0.13 \\
Bibiani-Anhwiaso-Bekwai    & 2111.42 & 209.24 & 0.10 &  Cape Coast                 & 1044.33 & 133.53 & 0.13 \\
Ga East                    & 465.00  & 47.28  & 0.10 &  Adansi South               & 586.25  & 80.29  & 0.14 \\
Atwima Mponua              & 986.75  & 101.07 & 0.10 &  Efutu                      & 1101.33 & 152.05 & 0.14 \\
Tarkwa-Nsuaem              & 1264.08 & 139.09 & 0.11 &  Asante Akim South          & 1856.92 & 261.11 & 0.14 \\
Juaboso                    & 1330.00 & 147.46 & 0.11 &  Aowin                      & 939.00  & 133.93 & 0.14 \\
\hline
\end{tabular}%
}
\end{table*}

\subsection{Posterior inference for temporal admission dynamics}\label{sec:bayes_inf}

\begin{figure*}
\begin{minipage}[H]{0.8\linewidth}
\centering
\includegraphics[width=\textwidth]{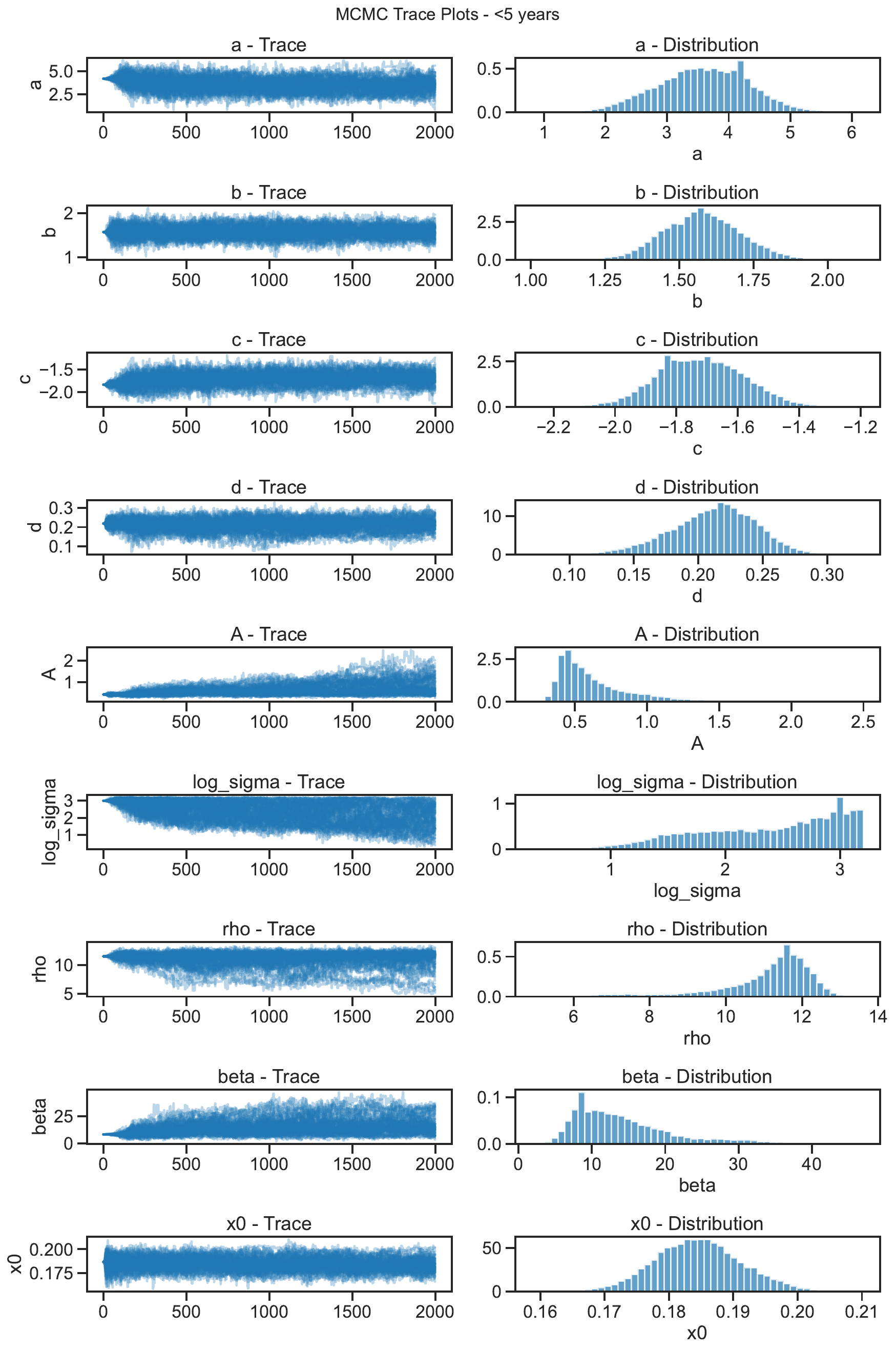} 
\end{minipage}
\caption{MCMC trace and posterior density plots for model parameters describing malaria admissions among children $< 5 $years in Ghana. Each parameter trace (n = 2000 samples) exhibits rapid convergence and good mixing, confirming sampler stability.}\label{fig:mcmc_trace_under5}
\end{figure*}

The MCMC trace plots for the $< 5$ years model (Figs.~\ref{fig:mcmc_trace_under5}) indicate well-mixed chains for the set of model parameters included in the hybrid cubic–DOK formulation.  
Across the sampled iterations the individual walker trajectories show no sustained drift or runaway excursions, and sample paths overlap substantially, which is consistent with adequate exploration of the posterior support.  
The marginal posterior densities plotted alongside the traces are unimodal for the majority of parameters and display credible mass concentrated within finite ranges; where posterior spreads are appreciable, the density shapes remain continuous and unimodal rather than multimodal.  
Taken together, these graphical diagnostics support the interpretation that the posterior for the $< 5$ years model is well characterised by the MCMC sample and that subsequent posterior summaries are reliable for inference and prediction (see also \citet{vehtari2021rank,foreman2013emcee} for convergence best practice).

% \citealp{foreman2013emcee}

\begin{figure*}
\begin{minipage}[H]{0.8\linewidth}
\centering
\includegraphics[width=\textwidth]{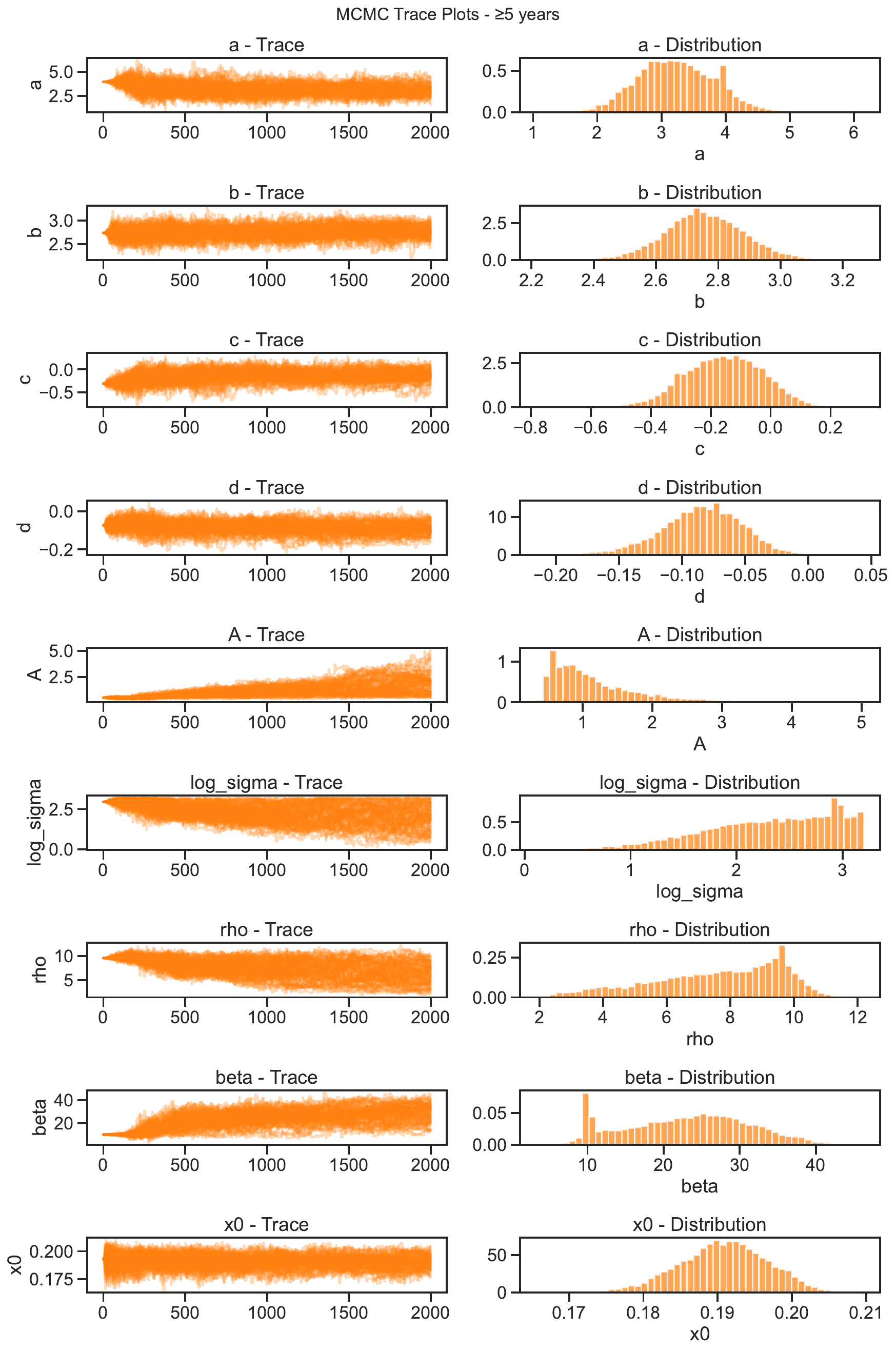} 
\end{minipage}
\caption{MCMC trace and posterior density plots for the $\geq 5$ years cohort. Chains display satisfactory convergence, though with broader posterior spreads reflecting greater inter-annual variability in older populations.}\label{fig:mcmc_trace_over5}
\end{figure*}

\begin{figure*}
\begin{minipage}[H]{\linewidth}
\centering
\includegraphics[width=\textwidth]{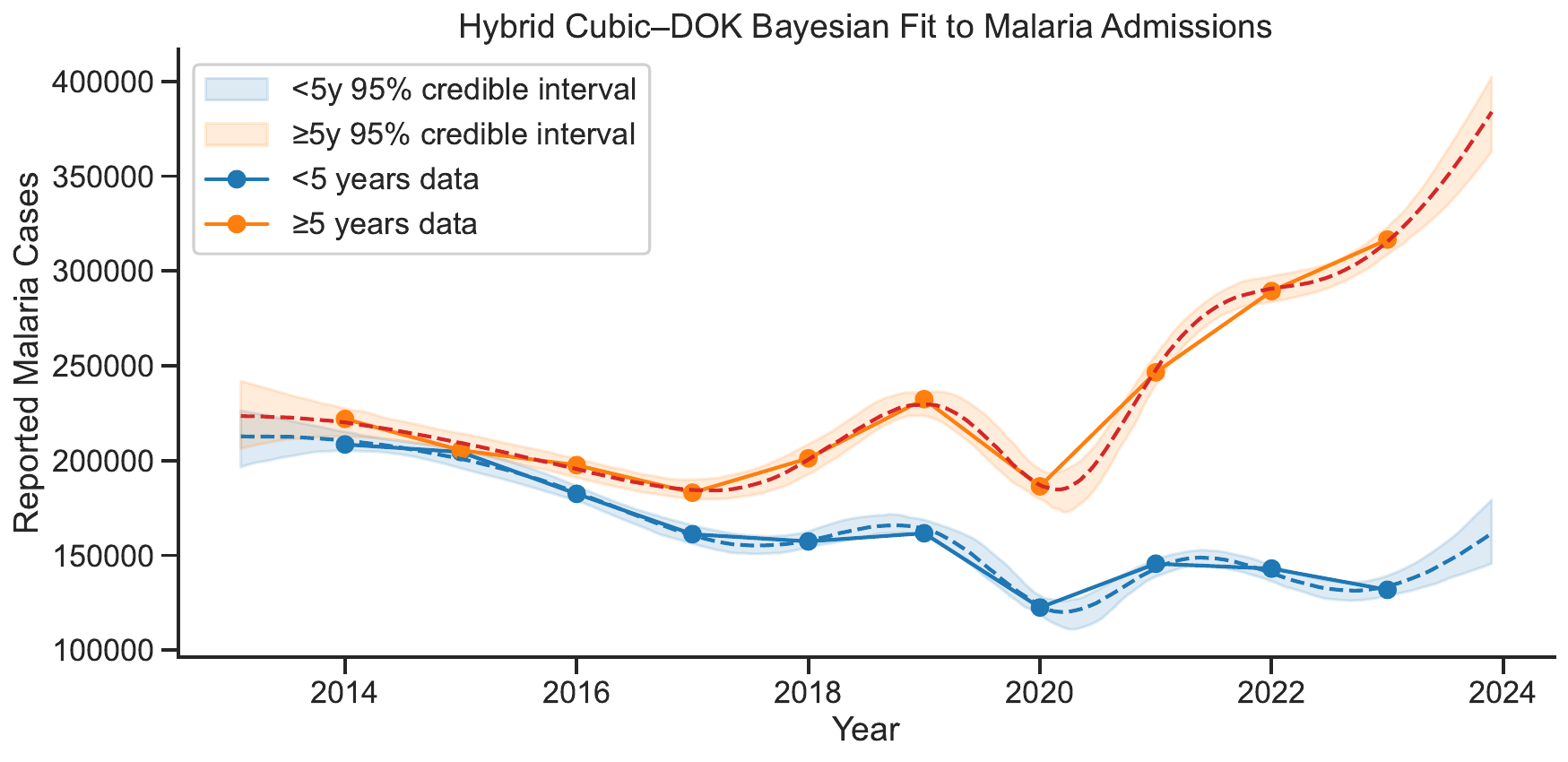} 
\end{minipage}
\caption{Hybrid Cubic–DOK Bayesian fit to annual malaria admissions in Ghana (2014--2023), with $95 \%$ credible intervals for both age groups.
Observed admissions (points) are well captured by the posterior median trajectories, demonstrating the model’s high predictive fidelity.}\label{fig:bayesian_posterior_fit}
\end{figure*}

The trace and density plots for the older cohort (\(\geq\)5 years) in Fig. \ref{fig:mcmc_trace_over5} show convergence characteristics broadly similar to the $< 5$ years model, with slightly wider marginal spreads for a subset of parameters.  
Chains reach a stationary appearance within the sampled iterations, and there is no visual evidence of persistent multimodality or non-mixing.  
The comparatively broader posterior densities for some parameters indicate greater uncertainty in those components of the mean function for the older cohort; this is visible as wider credible bands in the posterior predictive results (see Fig.~\ref{fig:bayesian_posterior_fit}).  
These diagnostic features are consistent with established practice for ensemble MCMC samplers \citep{foreman2013emcee, goodman2010ensemble} and argue for using full posterior predictive distributions rather than point estimates when making inference for the older cohort.

Figure~\ref{fig:bayesian_posterior_fit} presents the posterior median fits and $95\%$ credible intervals for both age groups superimposed on the observed annual admission series.  
Because the chains for both models exhibit the convergent and unimodal behaviour described above, the credible envelopes plotted in this figure provide statistically coherent representations of parameter and predictive uncertainty.  
Where the credible band widens in later years, this behaviour corresponds to increased posterior uncertainty arising from limited information in the training window or larger residual variance in those years; where it tightens, the posterior indicates stronger constraint by the data.  
The diagnostics in Table~\ref{tab:prediction_accuracy_compact} reflect time-varying posterior uncertainty, underscoring that full posterior predictive evaluation offers a more reliable measure of performance than deterministic fits.

%%%%%%%%% Posterior fit %%%%%%%%%%%%%

\begin{table}[htbp]
\centering
\caption{Prediction accuracy summary for the Bayesian hybrid Cubic–DOK model ($< 5$ years cohort.
Residuals (Actual – Predicted) and percentage errors indicate uniformly small deviations ($< 2 \%$), confirming the model’s calibration adequacy and absence of systematic bias.}
\label{tab:prediction_accuracy_compact}
\begin{adjustbox}{max width=\textwidth}
\begin{tabular}{@{}l l S[table-format=6.0] S[table-format=6.0] S[table-format=-4.0] S[table-format=2.1]@{}}
\toprule
\textbf{Year} & \textbf{Group} & \textbf{Actual} & \textbf{Predicted} & \textbf{Residual} & \textbf{Error (\%)} \\
\midrule
2014 & <5 years & 208499 & 209699 & -1200 & 0.6 \\
2015 & <5 years & 204498 & 202759 & 1739 & 0.9 \\
2016 & <5 years & 182453 & 182957 & -504 & 0.3 \\
2017 & <5 years & 161157 & 159163 & 1994 & 1.2 \\
2018 & <5 years & 157366 & 158024 & -658 & 0.4 \\
2019 & <5 years & 161657 & 164442 & -2785 & 1.7 \\
2020 & <5 years & 122478 & 123582 & -1104 & 0.9 \\
2021 & <5 years & 145587 & 143801 & 1786 & 1.2 \\
2022 & <5 years & 143022 & 140399 & 2623 & 1.8 \\
2023 & <5 years & 131811 & 133702 & -1891 & 1.4 \\
\bottomrule
\end{tabular}
\end{adjustbox}

\vspace{2pt}
\parbox{\textwidth}{\footnotesize \textit{Note:} Values rounded for compact presentation. Residual = Actual - Predicted.}
\end{table}

\subsection{Posterior predictive evaluation}
\label{sec:posterior_parameters}

\begin{figure*}
\begin{minipage}[H]{\linewidth}
\centering
\includegraphics[width=\textwidth]{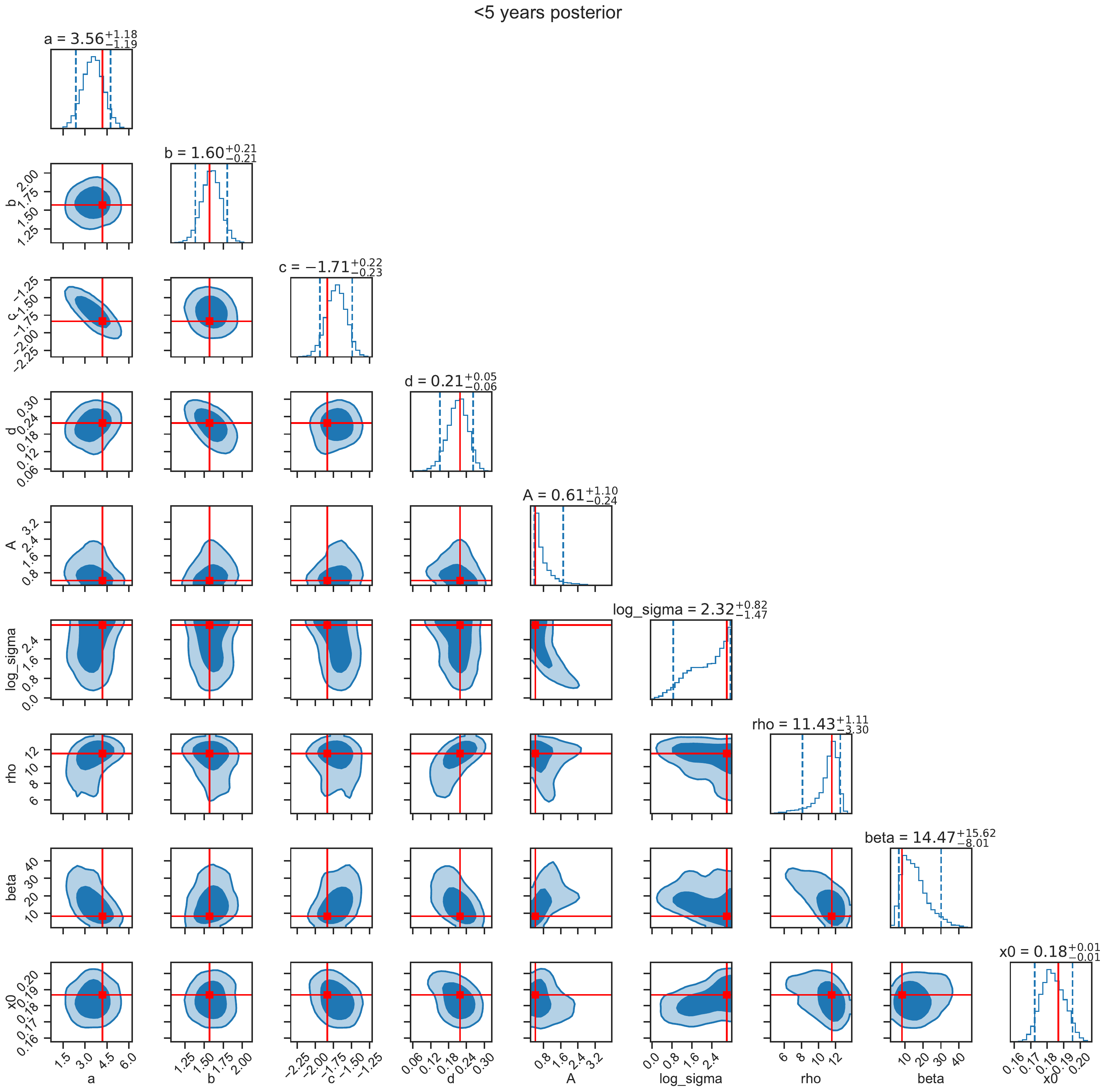} 
\end{minipage}
\caption{Joint and marginal posterior distributions for the hybrid cubic–DOK Bayesian model fitted to malaria admissions among children $<5$~years.  
Diagonal panels show marginal posteriors with median and 95\,\% credible intervals for each parameter; off-diagonal panels illustrate pairwise covariances.  
The unimodal and tightly constrained posteriors (\(a = 3.56^{+1.18}_{-1.19}\), \(b = 1.60^{+0.21}_{-0.21}\), \(c = 1.71^{+0.22}_{-0.23}\), \(d = 0.21^{+0.05}_{-0.06}\)) demonstrate strong parameter identifiability and confirm that the temporal trend and local oscillations are robustly estimated for the under-five cohort.}\label{fig:corner_plot_under5}
\end{figure*}

\begin{figure*}
\begin{minipage}[H]{\linewidth}
\centering
\includegraphics[width=\textwidth]{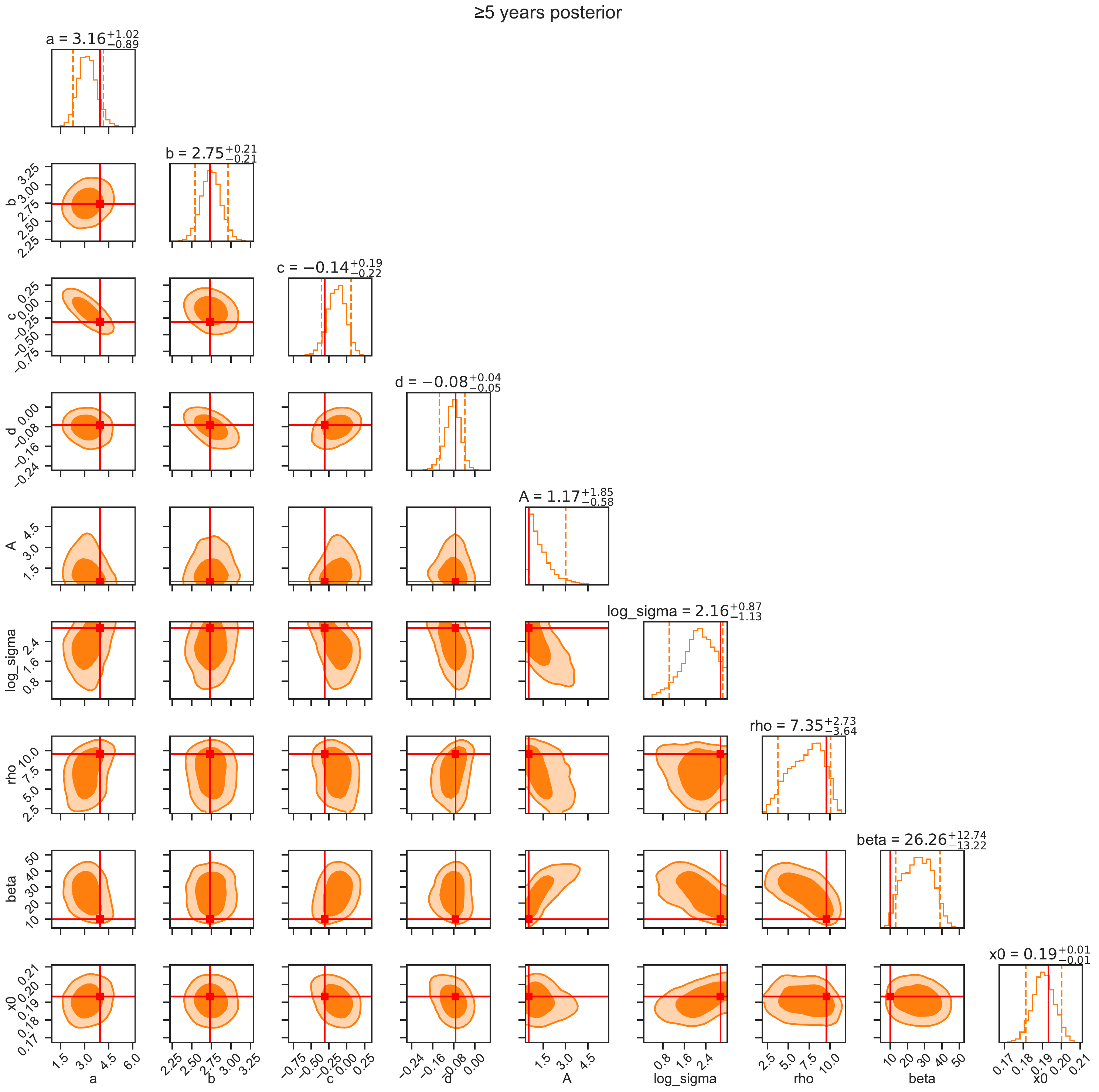} 
\end{minipage}
\caption{Posterior parameter correlations and marginal densities for the hybrid cubic–DOK Bayesian model fitted to malaria admissions among individuals aged~$\geq5$~years.  
Compared with the younger cohort, posterior spreads are broader for the kernel parameters (\(A = 1.17^{+1.85}_{-0.58}\), \(\rho = 7.35^{+2.73}_{-3.64}\), \(\beta = 26.26^{+12.74}_{-13.22}\)), reflecting greater temporal irregularity and higher uncertainty in short-term oscillations.  
Despite this, the posteriors remain unimodal and well mixed, confirming convergence and credible quantification of uncertainty in the adult time series.}\label{fig:corner_plot_over5}
\end{figure*}

Figures~\ref{fig:corner_plot_under5} and \ref{fig:corner_plot_over5} show the marginal posterior distributions and parameter covariances for the hybrid cubic–DOK model applied to the two age cohorts.  Each panel visualises the joint posterior space explored by the ensemble sampler, with diagonal entries representing marginal posteriors and off-diagonal cells depicting pairwise parameter correlations.  Overall, both models exhibit well-behaved, approximately symmetric posteriors with negligible multimodality, confirming that the likelihood surface is smooth and that the MCMC chains have converged to a stable region of high probability density.

For the younger cohort (in Fig.~\ref{fig:corner_plot_under5}), the cubic baseline parameters $(a,b,c,d)$ display sharply defined, unimodal posterior densities, each centred close to their respective means: 
$a = 3.56^{+1.18}_{-1.19}$, 
$b = 1.60^{+0.21}_{-0.21}$, 
$c = -1.71^{+0.22}_{-0.23}$, and 
$d = 0.21^{+0.05}_{-0.06}$.  
These values indicate a smooth, weakly nonlinear baseline with a modest downward curvature towards the later years of the study period (2014–2023).  
The parameters controlling the damped oscillatory kernel ($A$, $\sigma$, $\rho$, $\beta$, $x_0$) are more heterogeneous in scale and uncertainty, with 
$A = 0.61^{+1.10}_{-0.24}$, 
$\sigma = 2.32^{+0.82}_{-1.47}$, 
$\rho = 11.43^{+1.11}_{-3.30}$, 
$\beta = 14.47^{+15.62}_{-8.01}$, 
and 
$x_0 = 0.18^{+0.01}_{-0.01}$.  
In particular, the amplitude $A$ is positive in its posterior mode but shows asymmetric uncertainty (larger upper than lower bound), while the decay parameter $\beta$ is large on the normalised scale and exhibits broad credible limits; together these indicate transient, relatively sharp deviations from the baseline that are nevertheless constrained to attenuate. The concentration of $x_0$ near $0.18 (\pm 0.01)$ implies that the kernel is localised in time relative to the normalised index. 
These posterior structures are consistent with the high goodness-of-fit reported in Table~\ref{tab:fit_statistics}, where the posterior median fit attains $R^{2}\approx 0.9958$ for the $<5$ cohort, implying that the joint posterior explains the observed annual series closely while still admitting identifiable transient perturbations as quantified above.

The posterior marginals for the older cohort (Fig.~\ref{fig:corner_plot_over5}) share the same qualitative form but reveal systematically different scales and uncertainties. The polynomial baseline parameters are: 
$a = 3.16^{+1.02}_{-0.89}$, 
$b = 2.75^{+0.21}_{-0.21}$, 
$c = -0.14^{+0.19}_{-0.22}$ and 
$d = -0.08^{+0.04}_{-0.05}$. 
Relative to the younger cohort these coefficients retain pronounced curvature (positive cubic term), but the linear/quadratic combination indicates a different baseline shape (notably larger $b$ and smaller $c$ posterior means). Kernel parameters for the older group are estimated with larger absolute amplitude and wider tails: 
$A = 1.17^{+1.85}_{-0.58}$, 
$\log\sigma = 2.16^{+0.87}_{-1.13}$, 
$\rho = 7.35^{+2.73}_{-3.64}$, 
$\beta = 26.26^{+12.74}_{-13.22}$, 
and 
$x_0 = 0.19^{+0.01}_{-0.01}$. 
The larger median amplitude $A$ and wider credible interval reflect stronger and less regular transient departures from the baseline in the $\geq5$ data; the posterior for $\beta$ is also larger on average (and more uncertain), implying more rapid attenuation on the model’s normalised scale when surges occur, albeit with substantial uncertainty about the precise damping rate.
The covariance panels in Fig.~\ref{fig:corner_plot_over5} show mild dependencies (for example between $A$ and $\beta$), but the marginals remain unimodal and interpretable. The model fit statistics in Table~\ref{tab:fit_statistics} quantifies the empirical adequacy of these posteriors: the posterior median fit for the older cohort attains $R^{2}\approx 0.9956$, indicating comparable explanatory power to the $<5$ fit while accommodating the larger kernel-driven excursions noted above.

Given the short temporal length of the series, these in-sample fit metrics in Table~\ref{tab:fit_statistics} should be interpreted cautiously and are reported primarily as descriptive measures rather than evidence of generalisation performance.

%%%%%%%%% Posterior distribution %%%%%%%%%%%%%

%%%%%%

\begin{table*}[htbp]
\centering
\caption{Model Fit Statistics for Malaria Admission Predictions}
\label{tab:fit_statistics}
\begin{adjustbox}{max width=0.8\textwidth}
\begin{tabular}{@{}l S[table-format=1.5] S[table-format=1.5] S[table-format=-2.8] S[table-format=-2.8] S[table-format=1.8] S[table-format=1.5]@{}}
\toprule
\textbf{Age Group} & \textbf{$\chi^2$} & \textbf{Reduced $\chi^2$} & \textbf{AIC} & \textbf{BIC} & \textbf{$R^2$} & \textbf{RMSLE} \\
\midrule
<5 years & 0.00429 & 0.00429 & -59.54013 & -56.81686 & 0.99578 & 7.48534 \\
$\geq$5 years & 0.00445 & 0.00445 & -59.17965 & -56.45638 & 0.99558 & 7.94283 \\
\bottomrule
\end{tabular}
\end{adjustbox}
\end{table*}

\subsection{Forecasting and uncertainty-aware projections}
\label{sec:forecast_results}

\begin{figure*}
\begin{minipage}[H]{\linewidth}
\centering
\includegraphics[width=\textwidth]{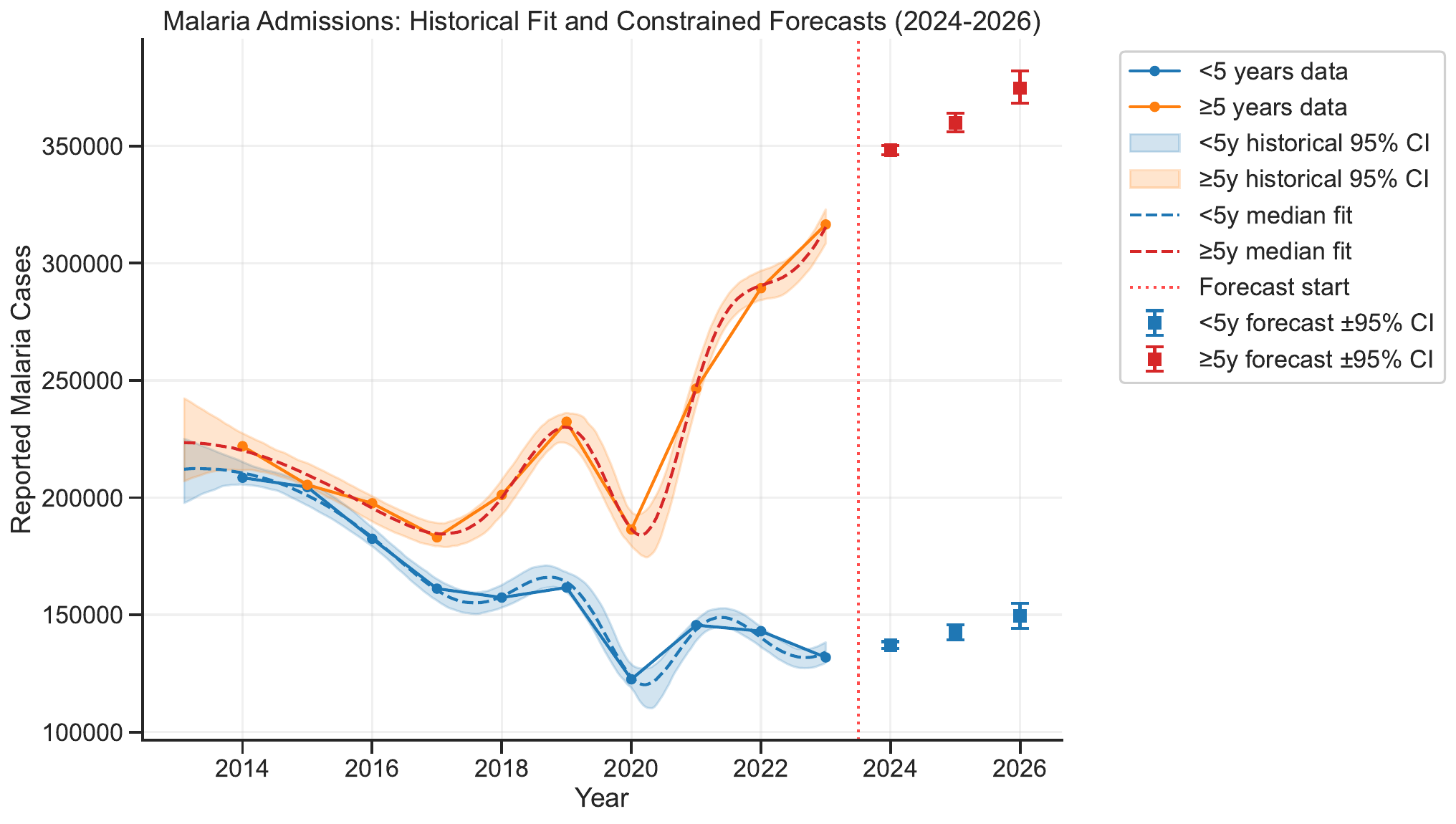} 
\end{minipage}
\caption{Posterior predictive forecasts of malaria admissions for 2024--2026 generated by the hybrid cubic–DOK Bayesian model.  
Shaded bands denote $95\,\%$ credible intervals derived from posterior propagation of parameter uncertainty.  
Both age groups show moderate upward trends with expanding uncertainty envelopes, reflecting credible accumulation of variance beyond the training period.}\label{fig:constrained_forecast_with_confidence}
\end{figure*}

Figure~\ref{fig:constrained_forecast_with_confidence} illustrates the three-year posterior predictive forecasts (2024--2026) derived from the hybrid cubic–DOK Bayesian models for both age groups, with 95\,\% credible intervals (CIs) reflecting parameter and observation uncertainty. 
These forecasts, summarised numerically in Table~\ref{tab:constrained_forecasts}, represent constrained posterior predictions obtained under the same normalisation and scaling conventions as the historical fits. 
The credible envelopes widen gradually with the forecast horizon, reflecting the expected accumulation of uncertainty beyond the training window.  

For children $< 5$ years, the model projects a gradual rise in malaria admissions from a posterior median of approximately $137{,}096$ (95\,\%\,CI: $135{,}658$--$138{,}547$) in 2024 to $149{,}500$ ($144{,}290$--$154{,}865$) by 2026. 
The corresponding year-on-year (YoY) increases of $+4.0$\,\%, $+3.9$\,\%, and $+4.9$\,\% indicate a sustained upward trajectory, albeit with modest growth and widening uncertainty bands (from 2.1\,\% to 7.1\,\% CI width). 
This pattern is consistent with the positive posterior means of the cubic coefficients $(a,b,c)$ reported earlier, which encode a mild long-term upward curvature superimposed on transient oscillatory deviations. 
The gradual broadening of the credible intervals towards 2026 is statistically coherent with the variance inflation expected under posterior propagation, where the cumulative predictive variance scales with both process uncertainty and finite-sample effects in the training data.  

Among individuals aged $\geq 5$ years, the model similarly anticipates continued increases, though from a substantially higher baseline. 
The median posterior forecasts rise from $348{,}377$ ($346{,}273$--$350{,}395$) in 2024 to $374{,}752$ ($368{,}357$--$381{,}997$) in 2026, corresponding to YoY increases of $+10.1$\,\%, $+3.3$\,\%, and $+4.1$\,\%, respectively. 
These growth rates are higher in the initial forecast year but stabilise subsequently, suggesting that much of the early rise reflects a return towards pre-2020 admission levels rather than an acceleration in underlying transmission intensity. 
The narrower credible ranges for this cohort (1.2--3.6\,\%) compared with those of the under-five group imply stronger posterior constraint and lower relative uncertainty, likely due to higher case counts and improved identifiability of the oscillatory kernel parameters ($A$, $\rho$, $\beta$).  

Notably, Fig.~\ref{fig:constrained_forecast_with_confidence} demonstrates that the hybrid cubic–DOK framework yields statistically coherent forecasts that balance long-term structural trends with bounded stochastic variation. 
The probabilistic forecasts remain internally consistent with the posterior distributions discussed in Figs.~\ref{fig:corner_plot_under5} and~\ref{fig:corner_plot_over5}, and quantitatively corroborated by the credible intervals in Table~\ref{tab:constrained_forecasts}. 
Such Bayesian predictive envelopes provide a more realistic depiction of uncertainty than point estimates alone, ensuring that future surveillance planning and intervention design are informed by credible probabilistic bounds rather than deterministic extrapolations \citep[e.g.][]{marwala2023hamiltonian,betancourt2017conceptual,vehtari2017practical,gelman2013bayesian}. 
In epidemiological terms, the observed widening of uncertainty reflects both stochastic variability in reporting and potential changes in transmission dynamics under shifting intervention coverage, reinforcing the utility of fully Bayesian temporal modelling for malaria burden forecasting in Ghana.

%%%%%%%%% Forecast distribution %%%%%%%%%%%%%

\begin{table*}[htbp]
\centering
\caption{Constrained Malaria Case Forecasts with Uncertainty Intervals (2024-2026)}
\label{tab:constrained_forecasts}
\begin{adjustbox}{max width=\textwidth}
\begin{tabular}{@{}l l S[table-format=6.0] c c c@{}}
\toprule
\textbf{Year} & \textbf{Age Group} & \textbf{Median Forecast} & \textbf{95\% CI Range} & \textbf{CI Width (\%)} & \textbf{YoY Change} \\
\midrule
2024 & <5 years & 137096 & 135658--138547 & 2.1 & +5285 (+4.0\%) \\
2025 & <5 years & 142492 & 139335--145780 & 4.5 & +5396 (+3.9\%) \\
2026 & <5 years & 149500 & 144290--154865 & 7.1 & +7008 (+4.9\%) \\
\addlinespace[0.2cm]
2024 & $\geq$5 years & 348377 & 346273--350395 & 1.2 & +31827 (+10.1\%) \\
2025 & $\geq$5 years & 360016 & 356150--364151 & 2.2 & +11639 (+3.3\%) \\
2026 & $\geq$5 years & 374752 & 368357--381997 & 3.6 & +14736 (+4.1\%) \\
\bottomrule
\end{tabular}
\end{adjustbox}
\vspace{2pt}
\parbox{\textwidth}{\footnotesize \textit{Note:}
 Forecasts based on constrained Bayesian modelling with realistic trend constraints. CI = Credible Interval, YoY = Year-over-Year change from previous year. Uncertainty increases with forecast horizon as expected.
 }
\end{table*}

\subsection{Implications for probabilistic decision support}
\label{subsec:decision_support}

The results presented above are intended to inform probabilistic decision support rather than to provide deterministic predictions or prescriptive intervention rules. By explicitly characterising uncertainty in both inferred temporal trends and short-term forecasts, the proposed framework supports expert judgement under conditions of limited data, reporting variability, and evolving epidemiological contexts. In this sense, the value of the model lies not in point forecasts alone, but in the structured quantification of plausible future trajectories and their associated uncertainty.

The posterior predictive distributions and forecast intervals highlight that uncertainty increases rapidly beyond the observed period, reflecting both parameter uncertainty and the limited temporal depth of the surveillance record. For planning and resource allocation contexts, such information is critical: widening predictive intervals signal reduced confidence in precise estimates and emphasise the need for flexible, adaptive planning rather than reliance on single-valued projections. Conversely, periods or age cohorts exhibiting comparatively narrower uncertainty bands may indicate more stable dynamics in reported admissions, supporting greater confidence in short-term expectations.

The age-specific contrasts observed in the posterior forecasts provide additional context for decision support by illustrating how uncertainty and temporal persistence differ across demographic groups. Rather than asserting shifts in underlying transmission dynamics, these contrasts can be interpreted as differences in the stability and predictability of reported admissions, which may inform prioritisation of monitoring efforts or sensitivity to emerging deviations from expected patterns. Importantly, the probabilistic formulation allows deviations from forecasted ranges to be identified as statistically meaningful departures, rather than as artefacts of random fluctuation.

From an operational perspective, the framework is compatible with expert-led surveillance and planning workflows in which probabilistic outputs are used to support situational awareness rather than to automate decisions. The model outputs can be interpreted alongside contextual information not explicitly modelled here such as intervention timing, health-system changes, or environmental anomalies--allowing domain experts to integrate quantitative uncertainty with qualitative knowledge. In this way, the framework functions as a transparent analytical layer that augments, rather than replaces, existing decision-making processes.

Overall, the proposed approach demonstrates how routinely collected malaria admission data can be transformed into uncertainty-aware summaries that support anticipatory reasoning under data constraints. By emphasising probabilistic interpretation and clearly communicating the limits of forecast confidence, the framework aligns with the needs of decision support in public-health settings where uncertainty is inherent and must be explicitly managed rather than ignored.

\section{Conclusion}
\label{sec:conc}

This study developed and demonstrated a probabilistic, uncertainty-aware Bayesian decision-support framework for analysing and forecasting malaria admissions under the severe data constraints characteristic of routine public-health surveillance systems. Rather than proposing a new class of artificial intelligence or mechanistic transmission models, the contribution lies in the design and validation of an interpretable Bayesian nonlinear inference pipeline that remains statistically identifiable, stable, and operationally meaningful when only short annual time series are available.

Methodologically, the framework integrates a low-dimensional cubic baseline with a bounded damped oscillatory kernel, estimated using an affine-invariant ensemble Markov chain Monte Carlo sampler. This hybrid cubic–DOK structure enables the separation of smooth long-term temporal evolution from transient, non-periodic deviations while avoiding the overparameterisation and instability that often accompany high-capacity machine-learning or stochastic state-space models in low-sample regimes. The probabilistic formulation yields full posterior distributions for all parameters and forecasts, thereby supporting uncertainty-aware reasoning rather than reliance on point estimates.

Applied to malaria admission data in Ghana for the period 2014–2023, the framework demonstrates strong inferential adequacy and coherent uncertainty quantification. Posterior diagnostics indicate stable convergence and well-mixed chains across a nine-dimensional parameter space, with high explanatory power for both age cohorts ($R^{2}=0.9958$ for children under five years and $R^{2}=0.9956$ for individuals aged $\geq 5$ years) and low residual error. Importantly, these in-sample fit statistics are interpreted descriptively rather than as evidence of generalisation, with model adequacy instead assessed through posterior predictive behaviour and uncertainty propagation.

The resulting posterior inference reveals distinct age-specific temporal patterns and pronounced spatial heterogeneity in reported malaria admissions. Admissions among children under five declined by approximately $35\%$ over the study period, while the older age cohort exhibited greater volatility and persistence. Spatially, coefficients of variation ranged from below $0.07$ in stable metropolitan districts such as Kumasi to above $3.3$ in districts such as Mpohor and Bia~East, reflecting heterogeneous reporting stability and surveillance fidelity. These findings illustrate how the proposed framework can extract interpretable signals from noisy, aggregated surveillance data without imposing strong mechanistic assumptions.

From a decision-support perspective, the principal value of the framework lies in its ability to generate probabilistic forecasts with explicit uncertainty bounds. Forecasts for 2024–2026 indicate a gradual resurgence in admissions for both age cohorts, accompanied by widening predictive intervals with increasing forecast horizon. This behaviour reflects principled uncertainty propagation rather than deterministic extrapolation and is essential for informing expert judgement under uncertainty. In this context, \enquote*{decision support} is understood as the provision of statistically defensible probabilistic evidence to inform planning discussions, resource prioritisation, and risk awareness, rather than the automation of allocation rules or policy decisions.

The study also clarifies the scope and limitations of the proposed approach. The framework is not intended to replace mechanistic transmission models, high-resolution spatio-temporal systems, or data-intensive machine-learning architectures when richer data are available. Nor does it explicitly model environmental drivers, intervention coverage, or reporting biases. Instead, it provides a robust analytical baseline for routine surveillance data, where limited temporal depth, aggregation effects, and reporting heterogeneity preclude more complex modelling strategies. By explicitly acknowledging these constraints, the framework promotes transparent and responsible use of statistical inference in applied decision-support settings.

Future extensions of this work include the incorporation of environmental and intervention covariates, hierarchical Bayesian structures for inter-district information sharing, and computationally efficient variational or sparse sampling schemes to support national-scale deployment. Embedding the framework within operational analytics platforms, such as early-warning dashboards or logistics planning tools, would enable continuous updating and iterative expert interpretation as new data becomes available.

In summary, this study demonstrates that rigorous Bayesian inference can serve as an effective expert decision-support component for public-health surveillance under realistic data limitations. By prioritising interpretability, uncertainty quantification, and methodological transparency, the proposed framework aligns with the aims of expert and intelligent systems that support informed human decision-making. Although illustrated through malaria admissions in Ghana, the approach is broadly applicable to other health and risk-monitoring contexts characterised by sparse, aggregated, and uncertainty-prone time series.

\section*{CRediT authorship contribution statement}

\textbf{T. Ansah-Narh:} Conceptualisation, Formal analysis, Validation, Methodology, Writing – original draft, and Writing – review and editing. \\

\textbf{Y. Asare Afrane:} Data curation, Funding acquisition, Project administration, Supervision, Resources, and Writing – review and editing. \\

\textbf{J. Bremang Tandoh:} Investigation, Visualisation, and Writing – review and editing.

\section*{Declaration of competing interest}
The authors declare that they have no financial or personal relationships that could be construed as potential conflicts of interest influencing the outcomes of this study.

\section*{Data availability} 
The dataset employed in this study consists of monthly, district-level malaria admission records for children under five years of age in Ghana, sourced from the Disease Surveillance Department of the Ghana Health Service. Owing to institutional confidentiality and data protection agreements, these data are not publicly accessible. However, access may be granted upon reasonable request to the Ghana Health Service, subject to the necessary ethical and administrative approvals. All analyses were performed on anonymised, aggregated datasets, thereby ensuring full compliance with data privacy standards and the exclusion of any patient-identifiable information.

\section*{Funding}
This work was supported by the Bill \& Melinda Gates Foundation (Grant No. INV-047051) through the West Africa Mathematical Modelling Capacity Development (WAMCAD) initiative, and by the National Institutes of Health (Grant No. D43 TW 011513).

% \section*{Acknowledgements}
% The authors gratefully acknowledge the West Africa Mathematical Modelling Capacity Development (WAMCAD) programme for its institutional and logistical support throughout this study. The WAMCAD initiative plays a pivotal role in strengthening regional expertise in mathematical and computational modelling of malaria and other neglected tropical diseases.
% The authors also express their appreciation to the Ghana Health Service for granting access to malaria surveillance data and providing technical insights that informed the analyses. TA-N acknowledges the computational resources provided by the High-Performance Computing (HPC) facility at the Ghana Space Science and Technology Institute, which enabled the model simulations and parameter inference.
% Finally, the authors thank the anonymous reviewers for their constructive feedback, which greatly improved the clarity, methodological rigour, and overall quality of the paper.

% \appendix
% \section{Example Appendix Section}


% %% For citations use: 
% %%       \citep{<label>} ==> [1]

% %%
% Example citation, See \citep{lamport94}.

 % \bibliographystyle{elsarticle-num-names.bst} 
  \bibliographystyle{model5-names.bst} 
 \bibliography{arefs}

\end{document}